\documentclass[%
reprint,
superscriptaddress,
amsmath,
amssymb,
aps,
prx,
floatfix,
]{revtex4-2}
\usepackage{graphicx}% Include figure files
\usepackage{dcolumn}% Align table columns on decimal point
\usepackage{bm}% bold math
\usepackage{hyperref}% add hypertext capabilities
\usepackage{xr}
\usepackage{lipsum}
\usepackage{braket}
\usepackage{xcolor}
\newcommand{\corr}[1]{\textcolor{black}{#1}}

\begin{document}
\preprint{APS/123-QED}
\bibliographystyle{unsrtnat} 
%=======================================================================================================================================
\title{Strong coupling between a microwave photon and a singlet-triplet qubit}
%%%%%
\author{J.\,H.~Ungerer}
\altaffiliation{Equal contributions.}
%\email{jannhinnerk.ungerer@unibas.ch}
\affiliation{
Department of Physics, University of Basel, Klingelbergstrasse 82 CH-4056, Switzerland
}
\affiliation{
Swiss Nanoscience Institute, University of Basel, Klingelbergstrasse 82 CH-4056, Switzerland
}
\author{A.~Pally\footnotemark[1]}
\altaffiliation{Equal contributions.}
\affiliation{
Department of Physics, University of Basel, Klingelbergstrasse 82 CH-4056, Switzerland
}
\author{A.~Kononov}
\affiliation{
Department of Physics, University of Basel, Klingelbergstrasse 82 CH-4056, Switzerland
}

\author{S.~Lehmann}
\affiliation{Solid State Physics and NanoLund, Lund University, Box 118, S-22100 Lund, Sweden}

\author{J.~Ridderbos}
\altaffiliation{Current address: MESA Institute for Nanotechnology, University of Twente, P.O. Box 217, 7500 AE Enschede, The
Netherlands}
\affiliation{
Department of Physics, University of Basel, Klingelbergstrasse 82 CH-4056, Switzerland
}

\author{P.\,P. Potts}
\affiliation{
Department of Physics, University of Basel, Klingelbergstrasse 82 CH-4056, Switzerland
}

\author{C.~Thelander}
\affiliation{Solid State Physics and NanoLund, Lund University, Box 118, S-22100 Lund, Sweden}
\author{K.A.~Dick}
\affiliation{Centre for Analysis and Synthesis, Lund University, Box 124, S-22100 Lund, Sweden}
\author{V.F.~Maisi}
\affiliation{Solid State Physics and NanoLund, Lund University, Box 118, S-22100 Lund, Sweden}
\author{P.~Scarlino}
\affiliation{Institute of Physics and Center for Quantum Science and Engineering, Ecole Polytechnique Fédérale de Lausanne, CH-1015 Lausanne, Switzerland}
\author{A.~Baumgartner}
\homepage{www.nanoelectronics.unibas.ch}
\affiliation{
Department of Physics, University of Basel, Klingelbergstrasse 82 CH-4056, Switzerland
}
\affiliation{
Swiss Nanoscience Institute, University of Basel, Klingelbergstrasse 82 CH-4056, Switzerland
}
\author{C.~Sch{\"o}nenberger}
\homepage{www.nanoelectronics.unibas.ch}
\affiliation{
Department of Physics, University of Basel, Klingelbergstrasse 82 CH-4056, Switzerland
}
\affiliation{
Swiss Nanoscience Institute, University of Basel, Klingelbergstrasse 82 CH-4056, Switzerland
}
\date{\today}
%=======================================================================================================================================
% ABSTRACT
%=======================================================================================================================================

\begin{abstract}
Tremendous progress in few-qubit quantum processing has been achieved lately using superconducting resonators coupled to gate voltage defined quantum dots. While the strong coupling regime has been demonstrated recently for odd charge parity flopping mode spin qubits, first attempts towards coupling a resonator to even charge parity singlet-triplet spin qubits have resulted only in weak spin-photon coupling strengths. Here, we integrate a zincblende InAs nanowire double quantum dot with strong spin-orbit interaction in a magnetic-field resilient, high-quality resonator. In contrast to conventional strategies, the quantum confinement is achieved using deterministically grown wurtzite tunnel barriers without resorting to electrical gating. Our experiments on even charge parity states and at large magnetic fields, allow to identify the relevant spin states and to measure the spin decoherence rates and spin-photon coupling strengths. Most importantly, we find an anti-crossing between the resonator mode in the single photon limit and a singlet-triplet qubit with an electron spin-photon coupling strength of \corr{$g/2\pi = 139 \pm 4$ MHz. Combined with the resonator decay rate $\kappa/2\pi=19.8\pm0.2$\,MHz and the qubit dephasing rate $\gamma/2\pi=116\pm7$\,MHz, our system achieves the strong coupling regime in which the coherent coupling exceeds qubit and resonator linewidth. These results pave the way towards large-scale quantum system based on singlet-triplet qubits.}
\end{abstract}
\maketitle
%%%% INTRO %%%%
%\section{Introduction}
Spin qubits in semiconductors are promising candidates for scalable quantum information processing due to long coherence times and fast manipulation~\cite{hanson2007spins,zwanenburg2013silicon,vandersypen2017interfacing,chatterjee2021semiconductor}.
For the qubit readout, circuit quantum electrodynamics based on superconducting resonators \cite{childress2004mesoscopic}, allows a direct and fast measurement of qubit states and their dynamics~\cite{petersson2012circuit}.
Recently, resonators were used to achieve charge-photon~\cite{stockklauser2017strong,mi2017strong}, spin-photon~\cite{mi2018coherent,samkharadze2018strong,landig2018coherent} as well as coherent coupling of distant charge~\cite{van2018microwave} and spin qubits~\cite{borjans2020resonant,harvey2022coherent}, enabling coherent information exchange between distant qubits.
However, the small electric and magnetic moments of individual electrons require complicated device architectures such as micromagnets, and a large number of surface gates that render scaling up to more complex architectures challenging. These approaches typically achieve a relatively weak electron spin-photon coupling on the order of $\sim 10 - 30$\,MHz.
In addition to single electron spin qubits, also spin qubits based on two electrons in a double quantum dot (DQD), e.g. in a singlet-triplet qubit have been demonstrated~\cite{petta2005coherent}.
Spin qubits based on two electrons typically offer a large hybridization of the spin and charge degree of freedom compared to single-electron spin qubits in principle allowing even stronger coupling strengths.
So far, however, the experimentally achieved coupling strengths in such systems~\cite{landig2019microwave,bottcher2022parametric} remained well below the strong coupling limit in which the coherent coupling rate exceeds both, the cavity mode decay rate and the qubit linewidth.

Here, we demonstrate that the strong coupling regime between a singlet-triplet qubit and a \corr{single photon in a} superconducting resonator can be reached.
We achieve this strong coupling by carefully designing the resonator and by using a DQD defined by in-situ grown tunnel barriers in a semiconductor with a large spin-orbit interaction.
The tunnel barriers consist of InAs segments in the wurtzite crystal-phase with an atomically sharp interface to the zincblende bulk of the nanowire (NW)~\cite{lehmann2013general}.
\corr{These crystal-phase barriers are highly reproducible and render the need of barrier gates obsolete, simplifying integration with superconducting resonators and making the nanowires a viable prototype for scalable quantum computing architectures.}

In this work, we make use of the large spin-orbit interaction in these nanowires~\cite{nilsson2018tuning} to define a singlet-triplet qubit at a finite \corr{in-plane} magnetic field in which the $T_{1,1}^+$ and $S_{2,0}$ states hybridize, forming a quantum two-level system.
Incorporating \corr{a NW with} a magnetic-field resilient resonator based on NbTiN~\cite{samkharadze2016high,ungerer2023performance} allows us to measure an avoided crossing between the singlet-triplet qubit and a single-photon excitation of the resonator at a \corr{ magnetic-field strength of $B=300$\,mT.
The measured coupling strength is very large compared to previously reported electron spin-photon coupling~\cite{mi2018coherent,samkharadze2018strong,landig2018coherent}, which enables us to reach the strong coupling regime.}
\corr{In addition, by analyzing the response of the hybridized resonator-qubit system for varying magnetic-field strengths, we perform qubit spectroscopy~\cite{borjans2021probing,mi2017high,ibberson2021large}.
This allows us to identify the specific spin states and to quantitatively extract the relevant device properties.}
\section{Device characterization}
\begin{figure}
    \centering
    \includegraphics[width=\linewidth]{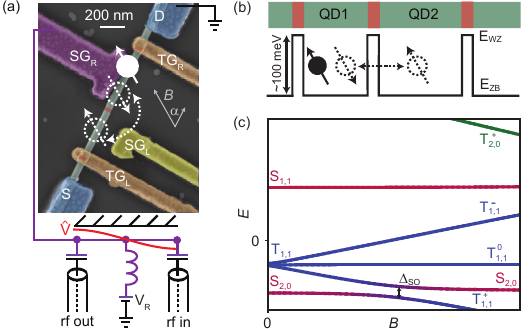}
    \caption{\textbf{Coupled resonator-qubit system} (a) False colored SEM-image of \corr{device A}. The NW (green) is divided into two segments by an in-situ grown tunnel barrier (red), thus forming the DQD system. The NW ends are contacted by two Ti/Au contacts (S,D) and the NW segements can be electrically tuned by two Ti/Au sidegates SG$_R$ (purple) and SG$_L$ (yellow).
    A high-impedance, half-wave resonator is connected to SG$_R$. Top gates (orange) are kept at a constant voltage of \corr{$-0.28$~V}. The magnetic field is applied in-plane at \corr{an angle $\alpha$ with respect to the NW axis, as illustrated by the grey arrow.} The white arrows illustrate an even charge configuration with the two degenerate DQD states $T_{1,1}^+$ and $S_{2,0}$. (b) Schematic of the crystal-phase defined DQD. The conduction band of wurzite and zincblende are offset by $\sim 100$ meV, resulting in a tunnel barrier between the zincblende segments. The intrinsic spin-orbit interaction enables spin-rotating tunneling between these segments. (c) Energy levels of an even charge configuration as a function of magnetic field $B$ at a fixed positive detuning $\varepsilon$ between the dot levels. %of $\varepsilon=2/2\pi$ GHz.
    At finite magnetic fields, $T_{1,1}^+$ (blue) hybridizes with $S_{2,0}$ (red) defining a singlet-triplet qubit with an energy splitting given by the spin-orbit interaction strength $\Delta_\mathrm{SO}$.}
    \label{fig:Fig1}
\end{figure}
\corr{Details about the NW properties and their growth can be found in the supplementary. The resonator-qubit system  of device A is shown in Fig.~\ref{fig:Fig1}(a), including a false-colored SEM-image of the crystal-phase defined NW DQD. We report similar experiments for two devices, A and B, with B discussed in the SI material. They are measured in a dilution refrigerator with a base temperature of $70$~mK.} The DQD forms in the \corr{490}~nm and \corr{370}~nm long zincblende segments (green), separated by $30$\,nm long wurtzite (red) tunnel barriers with a conduction band offset of $\sim$100\,meV~\cite{nilsson2016single}, as illustrated in Fig.~\ref{fig:Fig1}(b).
A high-impedance, half-wave coplanar-waveguide resonator is capacitively coupled to the DQD at its voltage anti-node via a sidegate.
In addition, the same sidegate can be used to tune the DQD charge states using a dc voltage ($V_R$) applied at the resonator voltage node. The DQD state is probed by reading out the resonator rf-transmission.
We extract the bare resonance frequency of the resonator \corr{$\omega_0/2\pi=5.1705\pm0.0003$\,GHz} at zero magnetic field and the bare decay rate \corr{$\kappa|_{B=0}/2\pi=27.3\pm0.6$ MHz.} %\kappa/2\pi=19.8\pm0.6$ at 300 mT
The resonator design and fitting are described in detail in methods section~\ref{app:resonator} and \ref{app:IO}. %\corr{Device B is shown and discussed in the supplementary material. In the following we focus exclusively on device A.}

In the following, we prepare the DQD in an even charge configuration in the many-electron regime (see methods~\ref{app:evenodd}), described by a two-electron Hamiltonian given in methods section~\ref{app:Hamiltonian}.
Figure \ref{fig:Fig1}(c) shows the eigenvalues of this Hamiltonian as a function of external magnetic field $B$ at a fixed DQD detuning.
At zero magnetic field, the detuning renders the singlet $S_{2,0}$ the ground state, for which both electrons reside in the same dot.
Without spin-rotating tunneling, this, and the $S_{1,1}$ state, with the electrons distributed to different dots, form a charge qubit~\cite{vanderwiel2002}. The subscripts describe the dot electron occupation of the left and right dot, respectively.
By applying an external magnetic field, the Zeeman effect lowers the energy of the triplet $T_{1,1}^+$ state, that becomes the ground state for sufficiently high magnetic fields. \corr{In the presence of a spin-rotating tunneling $t=\Delta_\mathrm{SO}/2$ induced by the intrinsic spin-orbit interaction $\Delta_\mathrm{SO}$, the energy levels of the hybridized $S_{2,0}$ and $T_{1,1}^+$ states are split.
The two new eigenstates of the avoided crossing form a singlet-triplet qubit shown schematically in Figs.~\ref{fig:Fig1}(a) and (b).}
%However, the intrinsic spin-orbit interaction hybridizes the $S_{2,0}$ and $T_{1,1}^+$ states, with the two new eigenstates of the avoided crossing forming a singlet-triplet qubit shown schematically in Figs.~\ref{fig:Fig1}(a) and (b).

\begin{figure}
    \centering
    \includegraphics[width=\linewidth]{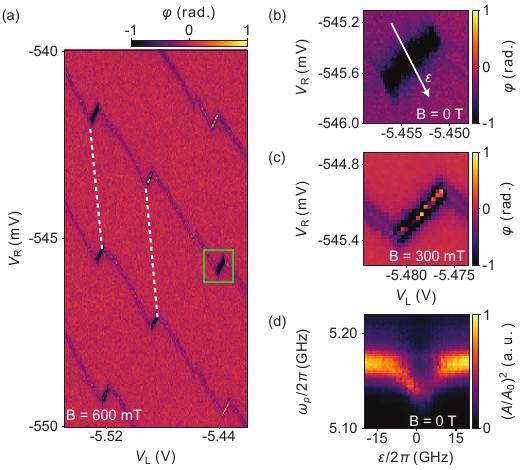}
    \caption{\textbf{Dispersive sensing of the DQD at $\mathbf{B=0}$.}
    (a) Charge stability diagram of the device \corr{at $B = 600$~mT applied at $\alpha=$ 164° with respect to the NW}, in which the resonator phase $\varphi$ is measured as a function of the SG voltages $V_R$ and $V_L$.
    %\corr{The negative slopes of the interdot transitions are due to the strong cross-capacitance of the larger gate $SG_R$}. 
    A zoom on the interdot transition pointed out by the \corr{green} rectangle is shown in (b) \corr{and (c) at $B = 0$~T and $B = 300$~mT with $\alpha= 57$°, respectively}. (d) \corr{Resonator transmission $(A/A_0)^2$} versus probe frequency $
    \omega_\mathrm{p}$ and detuning $\varepsilon$ (illustrated by the white line in (b)).
    At the charge degeneracy point of the DQD, \corr{we find a dispersive shift of $21\pm2$~MHz with respect to the bare resonance frequency. At small positive detuning a triplet state crosses the IDT, leading to a suppressed resonator transmission.}}
    \label{fig:Fig2}
\end{figure}
Figure \ref{fig:Fig2}(a) shows the charge stability diagram of \corr{device A} at a magnetic field of 600 mT with the angle $\alpha = 164$° with respect to the NW axis (See Fig.~\ref{fig:Fig1}(a)) detected as a shift in the transmission phase $\varphi$ of the resonator, plotted as a function of the two gate voltages $V_L$ and $V_R$ at a fixed probe frequency of \corr{$\omega_p/2\pi=5.174$ GHz}, close to resonance.
%i.e detuned from the bare resonace frequency by $80$ kHz.
We observe a \corr{characteristic honeycomb pattern of the charge stability diagram of a DQD.}
%The charge stability diagram of device B is discussed in S (ref).}
Using a capacitance model~\cite{van2002electron,scarlino2022situ}, we extract the gate-to-dot capacitances \corr{$C_{R2} = 44 \pm2$\,aF, $C_{L2} =2.0\pm0.2$\,aF, $C_{R1}=5\pm 2 $\,aF and $C_{L1} =4.6\pm0.2$\,aF for device A}.

\corr{We now focus on one particular inter-dot transition (IDT) marked by a green rectangle in Fig.~\ref{fig:Fig2}(a). The same IDT is shown in Fig.~\ref{fig:Fig2}(b) and (c) at $B=0$~T and $B= 300$~mT respectively, with $\alpha = 57$°. In Fig.~\ref{fig:Fig2}\corr{(d)} we show the normalized transmission $(A/A_0)^2$ at $B=0$~T,} while varying the probe frequency $\omega_p$ and relative detuning $\varepsilon_\mathrm{rel}$, \corr{illustrated by the} white line in Fig.~\ref{fig:Fig2}(b).
An electron can now reside on either of the two tunnel-coupled dots, constituting a charge qubit.
At the IDT, close to charge degeneracy, the electrical dipole moment of the charge qubit interacts with the resonator, resulting in a dispersive shift of the resonance frequency.
%Cut this part if necessary
By fitting \corr{input-output theory (see methods section~\ref{app:IO}) to this particular IDT, we extract the inter-dot tunnel coupling $t|_{B=0}/2\pi = 5.1\pm1.0$ GHz, the charge-photon coupling $g_0|_{B=0}/2\pi = 353\pm72$ MHz, and the charge qubit linewidth $\gamma|_{B=0}/2\pi = 1.7\pm0.7$~GHz.}
\section{Strong spin-photon coupling}
\label{sec:anti-crossing}
\begin{figure}[!htpb]
    \centering
    \includegraphics[width=\linewidth]{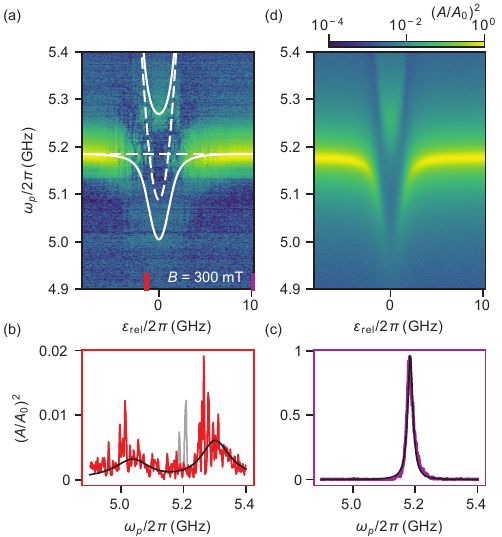}
    \caption{\textbf{Strong spin-photon coupling.} (a) \corr{Anti-crossing of the resonator and the qubit found when plotting the resonator transmission as a function of detuning $\varepsilon_\mathrm{rel}$ and probe frequency $\omega_p$ at a magnetic field of $B = 300$\,mT and $\alpha=57$°. The solid white curves are the eigenstate energies from fits to a Jaynes-Cummings model (Eq.~\eqref{eq:JComega} in methods~\ref{app:JC}).
    The faint double-peak structure at $\varepsilon\approx 0$ is an unambiguous signature of the strong coupling regime, $g>\kappa,\gamma$~\cite{blais2021circuit}.    
   (b,c) Cross sections at the detunings indicated by colored bars in (a).
   The solid lines stem from fit to input-output theory.
   (b) Double-peak structure at $\omega_q\sim\omega_0$ (see text). The larger noise floor for $\omega_p\sim\omega_0$ (grey data) is attributed to the bare resonator which is visible in spectroscopy because of a finite coupling between DQD and leads resulting in an odd DQD occupation for a short fraction of time during data acquisition.
   (c) Transmission for $\omega_q\gg \omega_0$, corresponding to the bare resonator.
   (d) Simulation using input-output theory with the parameters extracted from the input-output fit to (b).
   For these measurements, given the input-power $P_\mathrm{in}=-133$\,dBm, the average number of photons is $n<0.25$ (see methods section~\ref{app:Nphoton}).}}
    \label{fig:Fig3}\label{fig:Fig3review}
\end{figure}
When investigating the magnetic-field dependence of IDTs similar to the ones shown in  Fig.~\ref{fig:Fig2}(b,c), we observe two qualitatively different behaviors which we identify as even and odd charge parity configurations described in methods section~\ref{app:evenodd}. 
In the following, we investigate a single IDT, \corr{shown in Fig.~\ref{fig:Fig2}(c)}, with an even charge parity.

\corr{As illustrated in Fig.~\ref{fig:Fig1}(c), the DQD can be operated as a singlet-triplet qubit when applying a magnetic field.
The qubit frequency $\omega_q$ can be brought into resonance with the cavity frequency $\omega_0$ at \corr{$B\gtrsim200$\,mT}, as discussed in more detail below. At the resonance condition ($\omega_q\sim\omega_0$), an anti-symmetric (bonding) and a symmetric (anti-bonding) qubit-photon superposition state are formed.
The corresponding resonances can spectroscopically be discriminated only if the splitting $2g$ between them is larger than the dressed states' linewidth $\gamma+\kappa/2$~\cite{blais2021circuit}. In particular, the hybrid system is considered strongly coupled if the qubit-photon coupling strength $g$ exceeds $\gamma$ and $\kappa$~\cite{blais2021circuit}.}

\corr{In Fig.~\ref{fig:Fig3review}(a), we plot a spectroscopic measurement of the resonator where the singlet-triplet qubit is tuned into resonance by applying an electrostatic detuning $\varepsilon_\mathrm{rel}$ relative to the configuration at which $S_{2,0}$ and $T_{1,1}^+$ would be fully degenerate in the absence of a a spin-rotating tunneling.}
%While in Fig.~\ref{fig:Fig3review}(a) we tune the qubit transition frequency using the gate potentials, it can also be tuned by changing the magnetic field strength as demonstrated in Fig.~S3 in the supplementary.}
\corr{Consistent with strong coupling, we observe an avoided crossing between the resonator and the qubit.
At the points where the bare qubit frequency $\omega_q$ and resonator frequency $\omega_0$ (dashed, white curves) are degenerate, instead of crossing, they anti-cross.
And in Fig.~\ref{fig:Fig3}(a), a faint double peak structure is visible at around $\varepsilon_\mathrm{rel}\sim 0$ as $2g>\kappa/2+\gamma$, signature of the strong coupling regime~\cite{blais2021circuit}.}

\corr{For a quantitative analysis, we fit Lorentzians to the transmission of
 each trace of constant $\varepsilon_\mathrm{rel}$, we extract the transition frequencies $\omega_{\pm}$ of the dressed states.
 These are fitted to the Jaynes-Cummings model (solid, white curves in Fig.~\ref{fig:Fig3}(a)) described in methods section~\ref{app:JC}.
From this fit, we extract the tunnel rate $t|_{B=300\,{\rm mT}}/2\pi=\Delta_\mathrm{so}|_{B=300\,{\rm mT}}/4\pi=2.54\pm0.03$\,GHz and bare spin-photon coupling strength $g_{0}^\mathrm{JC}|_{B=300\,{\rm mT}}/2\pi = 123\pm16$ MHz. The extracted tunnel rate allows to plot the qubit transition frequency $\omega_q=\sqrt{(\Delta_\mathrm{so}/\hbar)^2+(\varepsilon_\mathrm{rel})^2}$ in Fig.~\ref{fig:Fig3review}(a) and to identify the resonance condition $\omega_q=\omega_0$ at a small electrostatic detuning $\varepsilon_\mathrm{rel}/2\pi=\pm 1.0$\,GHz. We evaluate the effective coupling strength  $g=g_0\cdot2t/\omega_q$ at the finite detuning $\varepsilon_\mathrm{rel}/2\pi=- 1.0$\,GHz and obtain $g^\mathrm{JC}|_{\epsilon_\mathrm{rel}/2\pi=-1\mathrm{\,GHz, }}/2\pi = 121\pm16$~MHz, as the spin-photon coupling strength on resonance condition.}

\corr{In Fig.~\ref{fig:Fig3review}(b), we plot a line trace at this detuning value as indicated in Fig.~\ref{fig:Fig3review}(a). Despite the large noise, the double peak structure is also clearly visible and stands in stark contrast to the bare resonator transmission at large detuning (see corresponding linetrace in Fig.~\ref{fig:Fig3review}(c)).
Using Eq.~\eqref{eq:transmission} derived from input-output theory described in the supplementary, we fit these data at 300\,mT and extract the spin-photon coupling strength $g_{\varepsilon_\mathrm{rel}=-\mathrm{1\,GHz}}/2\pi=139\pm4$\,MHz and qubit dephasing $\gamma/2\pi=116\pm7$\,MHz where we used the bare resonator decay $\kappa|_{B=300\sf\,mT}/2\pi=19.8\pm0.6$\,MHz. This value agrees well with the one obtained from the Jaynes-Cummings model. Using the values from input-output theory we model the whole anti-crossing using input-output theory in Fig.~\ref{fig:Fig3}(d), observing a very good agreement with the measurement.}

\corr{All together, this measurement therefore clearly demonstrates that the strong coupling regime between a single microwave photon and a singlet-triplet qubit is reached.}
\section{Magnetospectroscopy}
\begin{figure}
    \centering
    \includegraphics[width=\linewidth]{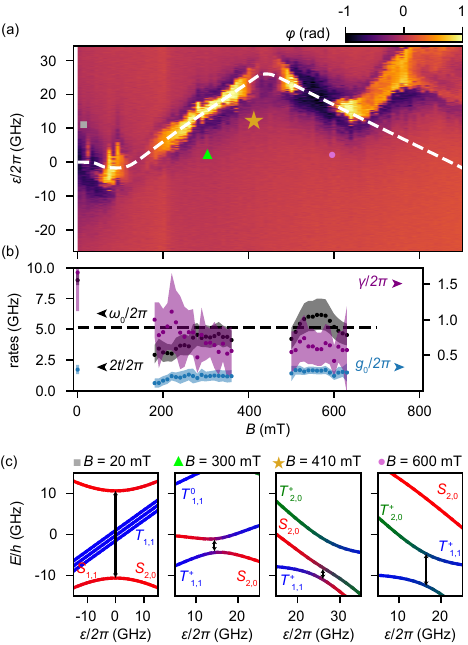}
    \caption{\textbf{Magnetospectroscopy of the singlet-triplet qubit.} a) Dispersive shift $\chi$ as a function of the magnetic field $B$ \corr{at an angle of $\alpha=130$°} and detuning $\varepsilon$. The white dashed line is a fit of the effective two-electron Hamiltonian (Eq.~\eqref{app:Hamiltonian}) to the data. b) Extracted tunnel rate $2t/2\pi$ (black), qubit-photon coupling $\corr{g_0}/2\pi$ (\corr{blue}) and qubit linewidth $\gamma/2\pi$ (\corr{purple}). The bare resonator frequency is indicated by the dashed black line. \corr{Shaded areas indicate the errorbars which originate from the uncertainty of the gate lever arm, which was independently measured.} (c) Two-electron energy level diagrams at various magnetic fields with the corresponding field strength indicated in (a) and (b) by the given symbols. For clarity a constant offset of 10 GHz, 20 GHz and 30 GHz was added to the energy levels at 300 mT, 410 mT and 600 mT. Given the input power $P_\mathrm{in}=-128$\,dBm, the average photon number is $n< 0.8$ in these experiments (see methods section~\ref{app:Nphoton}).}
    \label{fig:Fig4}
\end{figure}
\label{sec:magnetospectroscopy}
To explicitly identify and characterize the spin-orbit eigenstates and to independently verify the character of the singlet-triplet qubit, we now study the magnetic field evolution of the IDT from 0 up to \corr{$900$\,~mT applied at the angle $\alpha=130$°}.
%We measure the complex transmission amplitude  $S_{2,1}$ through the resonator as a function of magnetic field and DQD detuning. 
We measure the amplitude $A$ and phase $\varphi$ of the signal transmitted through the resonator as function of detuning $\varepsilon$ and magnetic field strength $B$.
A non-zero \corr{$\varphi$} occurs at the IDT when tunneling between the dots is allowed resulting in a non-zero DQD dipole moment.
As described in methods section~\ref{app:Hamiltonian}, we model the DQD by an effective two electron Hamiltonian which allows us to fit the gate voltage and field dependence of the IDT (white dashed line in \corr{Fig.~\ref{fig:Fig4}(a)}).
We find that the magneto-dispersion of the IDT is well described using the following fit parameters: the spin-conserving singlet and triplet tunnel rates \corr{$t_c^S/2\pi\approx8.5$\,GHz, and
$t_c^T/2\pi\approx3.2$\,GHz, the singlet-triplet coupling rate
$t_{\rm{SO}}/2\pi=\Delta_\mathrm{SO}/4\pi\approx 2.9$\,GHz, the electron g-factors of the right and left dots, $g_R\approx 1$ and $ g_L\approx 8$, as well as the single dot singlet-triplet energy splitting $\Delta_{\rm{ST}}/2\pi\approx 47$\,GHz.}
These fit parameters are consistent with parameters obtained previously in this material system~\cite{fasth2007direct,nadj2010disentangling,nilsson2018tuning,trif2008spin,junger2020magnetic}.
We note, however, that the fit is under-determined and therefore, it does not provide accurate numbers.
Nonetheless, the model gives a qualitative, physical understanding of the system and allows us to establish which DQD levels interact with the resonator.

Independently, we gain quantitative information about the system by considering the functional dependence of the amplitude $A$ and phase $\varphi$. % on the detuning $\varepsilon$. for values of fixed magnetic field $B$.
This is possible because the resonator provides an absolute energy scale allowing for a quantitative analysis of the interaction between the DQD and the resonator \corr{and hence to perform qubit spectroscopy~\cite{borjans2021probing,mi2017high,ibberson2021large}.
This spectroscopy complements} the preceding DQD Hamiltonian fit.
As described in methods \corr{section~\ref{app:IO}}, by fitting \corr{input-output theory to $\varphi$ and $A$} simultaneously, we extract the qubit tunnel amplitude $t$, the qubit linewidth $\gamma$, and the qubit-photon coupling strength $g$ as a function of $B$, which we plot in Fig.~\ref{fig:Fig4}(b). \corr{Here, we assume $\gamma$ as constant in detuning $\varepsilon$.}

Using the fits to both, the 2-electron Hamiltonian model and \corr{input-output theory} in the 2-level approximation, allows us to directly identify several regimes, in each of which the qubit has a different spin-character.
Fig.~\ref{fig:Fig4}(c) shows the corresponding DQD level structure based on the fit parameters as a function of $\varepsilon$ for different magnetic field.

At a low magnetic fields around \corr{$B = 20$\,mT}, the triplet states (blue curves) are Zeeman split and the ground-state curvature is dominated by the anti-crossing between $S_{1,1}$ and $S_{2,0}$ (red curves).
We find a singlet charge qubit in the weak coupling limit, i.e. the linewidth exceeds the charge-photon coupling by a factor of five.
The formation of an \corr{asymmetric double-dip} structure in $\varphi (\varepsilon)$ between \corr{$B\sim 0.01$\,T and $B\sim0.18$\,T} is explained by an interaction between the three states $S_{2,0}$, $S_{1,1}$ and $T_{1,1}^+$ as described in the supplementary material.
Traces of \corr{$\varphi(\varepsilon)$} with an \corr{asymmetric} double-dip structure cannot be described by \corr{a two-level input-output model} and are therefore not analysed quantitatively here. \corr{At $B\approx 50$\,mT, $\varphi$ becomes positive. Which we interpreted as a drop of the tunnel rate below the resonator frequency, $2t<\omega_0$}

As $B$ is increased, the triplet state $T_{1,1}^+$ becomes the ground state for $\varepsilon<0$, as shown in the second panel of Fig.~\ref{fig:Fig4}(c) for \corr{$B=300$\,mT.}
The spin-orbit interaction couples the singlet and triplet states, leading to an anti-crossing between $S_{2,0}$ and $T_{1,1}^+$, which constitutes a singlet-triplet qubit with $\omega_q=\Delta_\mathrm{SO}=2t_\mathrm{SO}$~\cite{mutter2021all,jirovec2021singlet}.
In this regime, at larger $B$, the resonance condition between $S_{2,0}$ and $T_{1,1}^+$ occurs at larger $\varepsilon$, because the energy of the bare $T_{1,1}^+$ state decreases with larger $B$ and the energy of $S_{2,0}$ decreases with larger $\varepsilon$.
Therefore, the IDT is observed at larger $\varepsilon$ for increasing $B$.

Consistent with the interpretation of the formation of a singlet-triplet qubit, we measure an approximately constant tunneling rate $t$ between \corr{$B\sim0.18$\,T and $B\sim0.36$\,T.}
In this regime, we extract the average spin-orbit tunneling rate to be \corr{$\bar{t}_{\rm{so}}=1.94\pm 0.02$\,GHz}. % where the error bar is the root variance.
%At $B\approx 1.3$\,T, $\chi$ becomes positive. This is interpreted as a drop of the tunnel rate below the resonator frequency, $2t<\omega_0$.
%This decline in $t$ is not captured by our simplified Hamiltonian model and we speculate that changes in the orbital structure of a many-electron DQD could be the reason.

\corr{At a magnetic field of \corr{$B\approx 370$\,mT,} the resonator phase $\varphi$ starts to vanish due to the the triplet state $T_{2,0}^+$ becoming relevant. The triplet state results in a level repulsion between $T_{2,0}^+$ and $T_{1,1}^+$ and hence leads to a reduced energy gap between the $S_{2,0}$ level and the $T_{1,1}^+$ level. In Fig.~\ref{fig:Fig4}(c), this is illustrated by the smaller energy gap (black arrow) at $B=410$\,mT compared to the one at $B=300$\,mT. Due to the The reduced energy gap, the resonator-qubit coupling on resonance ($\omega-q=\omega_0$ is and hence is the signal in  $\varphi$.}
 
The level structure at large magnetic fields is plotted exemplary for \corr{$B=600$\,mT} in the right panel of Fig.~\ref{fig:Fig4}(c).
In this regime, the ground-state of the DQD at the IDT is formed by a superposition of the $T_{2,0}^+$ and $T_{1,1}^+$ states.
%, forming a charge qubit with triplet spin character.
We find that the curve of Fig.~\ref{fig:Fig4}(a) turns back towards lower $\varepsilon$ for increasing $B$, which can be understood by noting that the spin-polarized triplets $T_{2,0}^+$ and $T_{1,1}^+$ form a charge qubit similar to the singlets at low field.
While the transition is increasingly dominated by the triplet-charge qubit for increasing $B$, $\varphi$ becomes \corr{negative} at the IDT, because the anti-crossing between the triplet states $T_{2,0}^+$ and $T_{1,1}^+$ occurs at much larger frequencies, \corr{$2t_c^T > 2t_\mathrm{SO},\omega_0$.
Hence, the triplet charge qubit frequency does not cross the resonator frequency, leading to a negative phase shift.}

\corr{At fields $B>700$~mT the dispersion turns to higher $\varepsilon$ again. Which is not accounted for in our model. A possible explanations to this discrepancy is that the magnetic field not only affects the detuning $\varepsilon$ of the DQD but also the total energy. This results in the lead to dot transitions starting to influece the IDT at high magnetic fields. Nevertheless, the data is well described at the magnetic field strengths we investigate in detail.}

This large number of detailed findings justify the parameters of the two-electron Hamiltonian introduced above, which, in turn, directly allows us to identify the singlet-triplet spin qubit, for which we find the strong coupling limit to the electromagnetic cavity. 
%We note that the qubit linewidth $\gamma$ and qubit-photon coupling strength are both related to the qubit rate.
%To confirm their linear relation, we plot $\gamma$ and $g$ against $t$ in Fig.~\ref{fig:A2} in the appendix. 
%An intuitive explanation is that the tunnel rate in our experiments increases as the qubit becomes more charge like and hence is more susceptible to charge noise.
%Another possible explanation is that the qubit linewidth is limited by qubit relaxation which scales proportional to the tunnel rate to the contacts. \corr{This argument is further strengthened by the comparison of device A and device B. For device A a more negative voltage ($V_{TG_A}-0.28$ V and $V_{TG_B}=-0.05$ V) was applied on the top gates on top of the tunnel barriers to the leads. Consequently, we observed a smaller qubit linewidth compared to device B (see supplementary information for the data on device B).

Note, that the extracted qubit linewidth is larger in Fig.~\ref{fig:Fig4}(b) compared to the strong-coupling in Fig.~\ref{fig:Fig3}. This is caused by applying the magnetic field at different angles in the two measurements.%~\cite{pally2023inprep}.
%%%%%%%%%%%%%%%
%%%%%%%%%%%%%%%
%%%%SUMMARY%%%%
%%%%%%%%%%%%%%%
%%%%%%%%%%%%%%%
\section{Conclusion and Outlook}
In summary, we demonstrate a semiconductor nanowire DQD device with crystal-phase defined tunnel barriers that can be operated as different types of qubits, coupled to a high-impedance, high magnetic field resilient electromagnetic resonator. As the main result, we find
a singlet-triplet qubit for which we extract the relevant qubit parameters, especially \corr{a large electron spin-photon coupling of $g/2\pi=139$\,MHz in the single photon limit, reaching the strong coupling regime $g>\gamma,\kappa$.}

Our experiments demonstrate that deterministically grown tunnel barriers allow for a reduced number of gate lines, and that, mediated by intrinsic spin-orbit interaction, singlet-triplet qubits can reach the strong coupling limit for low photon numbers, similar to flopping mode spin qubits~\cite{yu2023strong,burkard2021semiconductor}. This finding is potentially applicable to other promising platforms with strong spin-orbit interactions, like holes in Ge~\cite{jirovec2021singlet}.
Our nanowire platform without depletion gates results in a significantly reduced gate-induced photon-leakage in the absence of on-chip filtering~\cite{petersson2012circuit,mi2017circuit,harvey2020chip}. \corr{And, since DQD parameters (such as charging energy and individual tunnel rates) can be set deterministically in the NW growth, multiple NWs with optimal and essentially identical characteristics properties can be obtained simultaneously~\cite{nilsson2017parallel} and possibly integrated on the same substrate~\cite{ram2021high}.
This drastically simplifies the search for an optimal gate regime and renders further gates, such as the top gates in our device, unnecessary.}
An optimized gate design with resonators with larger impedance~\cite{scarlino2022situ} therefore presents an ideal platform to investigate new phenomena in the ultrastrong coupling regime~\cite{forn2019ultrastrong,scarlino2022situ}. Additionally, the large electron spin-photon coupling found in our experiments will be crucial for the implementation of two-qubit gates between distant spin qubits, a milestone on the way towards scalable quantum computers.
%\corr{In the future, a scalable quantum computing architecture based on crystal-phase defined barriers might be incorporated using template-assisted selective epitaxy (“TASE”) in which nanowires are used as crystal seeds for creating a two-dimensional crystal.}

We acknowledge fruitful discussions with Simon Zihlmann, Roy Haller, Andrea Hofmann, Stefano Bosco, Romain Maurand and Antti Ranni, and support in setting-up the experiments by Fabian Op\corr{p}liger, Roy Haller, Luk Yi Cheung, and Deepankar Sarmah.
This research was supported by the Swiss Nanoscience Institute (SNI), the Swiss National Science Foundation through grant 192027, the NCCR Quantum Science and Technology (NCCR-QSIT), the NCCR Spin Qubit in Silicon (NCCR-Spin) and the Eccellenza Professorial Fellowship PCEFP2\textunderscore194268. We further acknowledge funding from the European Union’s Horizon 2020 research and innovation programme, specifically the FET-open project AndQC, agreement No 828948 and the FET-open project TOPSQUAD, agreement No 847471. Furthermore, we acknowledge funding by NanoLund and the Knut \& Alice Wallenberg Foundation (KAW). PS acknowledges support from the SNSF through grant 200418 and the SERI through grant 589025. All data in this publication are available in
numerical form at: \url{https://doi.org/10.5281/zenodo.7777840}.
\section{Methods}
\subsection{Resonator characterization and analysis}
\label{app:resonator}
The resonator is fabricated from a thin-film NbTiN \corr{(thickness $\sim$10\,nm)}, sputtered onto a Si/SiO\textsubscript{2} (500\,\textmu m/100\,nm) substrate~\cite{ungerer2023performance}.
These resonators can be operated for in-plane fields exceeding $5$\,T~\cite{samkharadze2016high,ungerer2023performance}.
The large sheet kinetic inductance of the used NbTiN film of $L_\mathrm{sq}\approx 90$\,pH combined with the narrow center conductor width of $\sim380$\,nm, and the large distance to the ground plane of $\sim 35$\,\textmu m results in an impedance of $2.1$ k$\Omega$.
The resonator can be dc biased using a bias line which contains a meandered inductor ensuring sufficient frequency detuning between the half-wave resonance used in the experiment and a second, low quality resonance mode at a lower frequency that forms due to the finite inductance of the bias line~\cite{harvey2020chip}.
\corr{A scanning electron micrograph of the resonator center-conductor is shown in Fig.\ref{fig:FigA1}(b) in the extended data.}
One of the two resonator voltage anti-nodes is galvanically connected to gate SG$_R$ shown in Fig.~\ref{fig:Fig1}(c) of the main text.
\subsection{Charge parity determination\label{app:evenodd}}
We measure the phase $\varphi$ and amplitude 
$A$ of the resonator as a function of detuning $\varepsilon$ and magnetic field $B$ at a probe-frequency $\omega_{\rm{p}}/2\pi=5.253$\,GHz, close to the bare resonator frequency.
A change in $\varphi$ reflects the dispersive interaction between the resonator and two anticrossing levels of the DQD~\cite{frey2012dipole,crippa2019gate}.
Therefore, the non-zero phase response of the resonator tracks the position of the IDT along the detuning axis. 
The comparison of the magnetic field dependence of the IDT position to a Hamiltonian model of the DQD allows one to determine the charge parity~\cite{crippa2019gate,ezzouch2021dispersively}.
Figures \ref{fig:Fig2_app} (a) and (b) in the extended data show two typical low field IDT characteristics \corr{of device B}.

For an odd number of electrons (Fig.~\ref{fig:Fig2_app}(b)), the DQD resonance gate voltage $V_R$, at which the IDT is observed, disperses linearly with magnetic field starting from zero. 
This can be understood considering the Zeeman-splitting of the unpaired electron energy levels and two non-equal Landé g-factors of the two dots.
Fig.~\ref{fig:Fig2_app}(c) shows the energy level diagram of a one-electron Hamiltonian including Zeeman-splitting with a g-factor difference of 1.0 and spin-orbit interaction $t_{SO}/2\pi = 5$~GHz at a magnetic field of $B=500$\,mT (green, dashed line in Fig.~\ref{fig:Fig2_app}(b). The one-electron Hamiltonian is explicitly discussed in the supplementary material.
The arrow points out the center of the IDT (largest curvature of the groundstate~\cite{park2020adiabatic}) which corresponds to the largest dipole moment of the DQD and thus to the largest change in $\varphi$.
This point shifts with $B$ towards increasingly negative values.

For an even number of electrons in the DQD at zero magnetic field (Fig.~\ref{fig:Fig2_app}(a)), a single dip in phase is observed, but at a low magnetic fields, $B\approx60$\,mT, a double dip structure emerges as a function of $\varepsilon$ (see supplementary material for details).
This double-dip originates from an interaction between $S_{2,0}$, $S_{1,1}$ and $T_{1,1}^+$ as explained in detail in the supplementary material.
The dependence of the IDT on magnetic field for an even number of electrons can be understood using an effective two electron Hamiltonian including spin-orbit interaction described in more detail in section~\ref{app:Hamiltonian}.
In Fig.~\ref{fig:Fig2_app}(c), we plot the energy levels at a magnetic field $B = 0.15$~T.
In contrast to the odd filling, starting at zero magnetic field, the arrow marking the center of the IDT barely changes, consistent with our measurement. 
The double dip vanishes when further increasing the magnetic field, because of an increasing occupation of the polarized triplet states.
Once the Zeeman energy of the triplet state $\ket{T_{1,1}^+}$ becomes comparable to the singlet charge tunneling $t_{\rm{c}}^{\rm{S}}$, the position of the IDT as a function of $B$ disperses towards larger $\varepsilon$~\cite{schroer2012radio,malinowski2018spin,ezzouch2021dispersively}.
This transition is marked by the white dashed line at $0.2$~T in~\ref{fig:Fig2_app}(a).

Based on the good qualitative agreement between our data and the one electron and two electron Hamiltonian, respectively, we can clearly identify the even and odd charge parities.

\subsection{Jaynes-Cummings model}
\label{app:JC}
In the regime of only two DQD levels being relevant, we model the DQD Hamiltonian as an effective two-level system (qubit) interacting with a single \corr{mode} in the resonator.
The combined system is described by the Jaynes-Cummings model~\cite{shore1993jaynes} \corr{
\begin{equation}
    \label{eq:hamiltonJC}
    \hat{H}/\hbar = \omega_0\hat{a}^\dagger\hat{a}+\frac{\omega_q}{2}\hat{\sigma}_z+g\left(\hat{a}\hat{\sigma}^\dagger+\hat{a}^\dagger\hat{\sigma}\right),
\end{equation}
where $\hat{a}$ is the photon annihilation operator, $\hat{\sigma}$ the qubit lowering operator, and $\hat{\sigma}_z$ the Pauli z-matrix in the qubit subspace. The qubit frequency is given by $\omega_q=\sqrt{\left(2t\right)^2+\varepsilon^2}$~\cite{vanderwiel2002} with the effective qubit-photon coupling strength $g=g_0\cdot2t/\omega_q$ accounting for the mixing angle~\cite{blais2004cavity,stockklauser2017strong}, where $g_0$ is the bare qubit-photon coupling.}
\corr{An excitation from the ground state has the transition frequency}~\cite{blais2004cavity}
\begin{equation}
\omega_{\pm}=\corr{\frac{\omega_0+\omega_q}{2}}\pm\frac{1}{2}\sqrt{4g^2+(\omega_0-\omega_q)^2}\text{.}\label{eq:JComega}
\end{equation}

\subsection{Input-Output theory}
\label{app:IO}
\corr{To derive the response of the resonator, we use the equations of motion \cite{gardiner2004quantum}
\begin{equation}
    \label{eq:ainoutgen}
    \begin{aligned}
        &\partial_t{\langle \hat{a}\rangle}(t) = -i\omega_0 \hat{a}(t)-ig\langle \hat\sigma\rangle (t)-\frac{\kappa}{2}\langle \hat{a}\rangle(t)\\&\hspace{2cm}-\sqrt{\kappa_1}\langle \hat{b}_{{\rm in}, 1}\rangle(t)-\sqrt{\kappa_2}\langle \hat{b}_{{\rm in}, 2}\rangle(t),\\&
        \partial_t \langle \hat{\sigma}\rangle (t) = -i\omega_q\langle \hat{\sigma}\rangle (t)+ig\langle \hat{a} \hat{\sigma}_z\rangle(t)-\gamma\langle \hat{\sigma}\rangle (t).
    \end{aligned}
\end{equation}
%where $\kappa$ denotes the line-width of the cavity and $\gamma$ the dephasing rate of the DQD. 
The input couplings are denoted by $\kappa_{j}$ and the operators $\hat{b}_{{\rm in}, j}(t)$ capture a coherent drive in port $j$. In our experiments $\kappa_1\approx\kappa_2\approx\kappa/2$ as the resonator is symmetrically coupled and operates in the strongly over-coupled regime. The output of the cavity can be computed from the input-output relation \cite{gardiner2004quantum}
\begin{equation}
    \label{eq:inout}
    \langle \hat{b} _{{\rm out},j}\rangle(t) = \langle \hat{b}_{{\rm in},j}\rangle(t)+ \sqrt{\kappa_j}\langle \hat{a}\rangle(t).
\end{equation}
To solve these equations, we approximate \cite{wong:2017,schondorf:2018}
\begin{equation}
	\label{eq:mean_field}
	\langle\hat{a} \hat{\sigma}_z\rangle(t)\rightarrow \langle \hat{a}\rangle (t)\langle \hat{\sigma}_z\rangle,
\end{equation}
where $\langle \hat{\sigma}_z\rangle$ is evaluated at steady state and captures the difference between the population of the excited qubit state and the ground state, accounting for operation at larger temperatures or drive strengths. In our experiments, we operate in the linear regime, $\langle \hat{\sigma}_z\rangle= -1$.}

\corr{To compute the transmission amplitude, we solve Eqs.~\eqref{eq:ainoutgen} and \eqref{eq:inout} upon Fourier transformation and set $\langle \hat{b}_{{\rm in},2}\rangle(t)=0$. This results in the transmission amplitude
\begin{equation}
    \label{eq:transmission}
    \tau(\omega) = -\frac{\langle \hat{b} _{{\rm out},2}\rangle(\omega)}{\langle \hat{b} _{{\rm in},1}\rangle(\omega)}= \sqrt{\kappa_1\kappa_2}A(\omega),
\end{equation}
where the minus sign accounts for the phase difference of $\pi$ between the input and the output port ($\lambda/2$ resonator) and
\begin{equation}
    \label{eq:aomega}
    A(\omega) = \frac{\gamma+i(\omega_q-\omega)}{[\kappa/2+i(\omega_0-\omega)][\gamma+i(\omega_q-\omega)]-g^2\langle \hat{\sigma}_z\rangle}.
\end{equation}
In the main text, the absolute value squared of this quantity normalized by its maximal value is shown.}

\begin{figure}[htbp!]
    \centering
    \includegraphics[width=\linewidth]{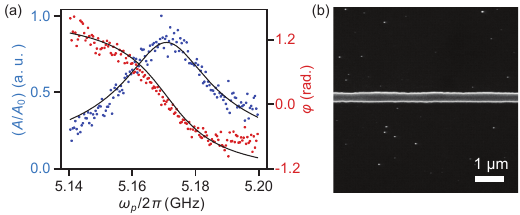}
    \caption{\textbf{Resonator of a device A.}
    \corr{(a) Resonance curve of the resonator in Coulomb blockade in amplitude $A/A_0$ (blue) and phase $\varphi$ (red). The black lines are simultaneous fits to the data using input-ouput theory. (b) Scanning electron micrograph of the resonator center conductor.}}
    \label{fig:FigA1}
\end{figure}

\begin{figure}[htbp!]
    \centering
    \includegraphics[width=\linewidth]{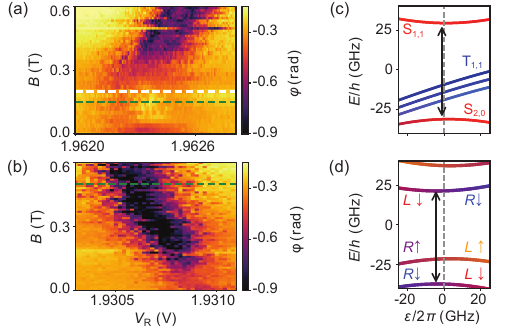}
    \caption{\textbf{Dispersive read-out at low magnetic field.}
    Resonator phase in dependence of the right gate voltage $V_R$ and magnetic field $B$ for even (a) and odd (b) occupation of the DQD \corr{of device B}. For the odd occupation the IDT shifts to lower $V_R$ from $B=0$. The IDT of the even occupation stays nearly independent of magnetic field until around $0.2$~T (white dashed line), from where it starts moving to more positive $V_R$. Energy level diagram for the even (c) and odd (d) configuration at $0.15$~T  and $0.5$~T (green dashed line). The arrow marks the transition the resonator is sensitive to, where the ground state energy level has maximum curvature.}
    \label{fig:Fig2_app}
\end{figure}

\begin{figure}[htbp!]
    \centering
    \includegraphics[width=\linewidth]{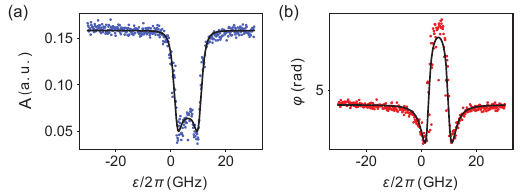}
    \caption{\textbf{Fit of input-output theory to the data.}
    \corr{Input-output theory (solid lines) simoultaneously fitted to the resonator amplitude $A$ (a) and the resonator phase $\varphi$ (b) as a function of detuning $\varepsilon$ of the even configuration at $0.25$ T.}}
    \label{fig:Fig_app_shift_linewidth}
\end{figure}

%\corr{In input-output theory~\cite{gardiner2004quantum}, as derived in the supplementary, the transmission amplitude of the coupled resonator-qubit hybrid system is given by}
%\begin{equation}\label{eq:A_io}
%A(\omega)=\frac{\sqrt{\kappa_L\kappa_R} \left[\gamma+i\left(\omega_q-\omega\right)\right]}{\left[\frac{\kappa}{2}+i\left(\omega_0-\omega\right)\right]\left[\gamma+i\left(\omega_q-%\omega\right)\right]-g^2m_z}.
%\end{equation}
%Here, $m_z=p_e-p_g$ is the difference between the population of the excited qubit state $p_e$ and the population of the ground state $p_g$ and in principle accounts for operation at larger temperatures or drive strengths. In our experiments, we operate in the linear regime, $m_z=-1$. Furthermore $\kappa_L$ and $\kappa_R$ are the coupling rates of the resonator to the left and right port. In our experiments $\kappa_L\approx\kappa_R\approx\kappa/2$ as the resonator is symmetrically coupled and operates in the strongly over-coupled regime.
The phase of the transmitted signal is given by \corr{
\begin{equation}
	\label{eq:transphasegen}
 \begin{aligned}
	&\varphi(\omega) = -\arctan(\Lambda),\\&\Lambda =\frac{-2(\omega_q-\omega)g^2\langle\hat{\sigma}_z\rangle-2(\omega_0-\omega)[\gamma^2+(\omega_q-\omega)^2]}{\kappa[\gamma^2+(\omega_q-\omega)^2]-2\gamma g^2\langle \hat{\sigma}_z\rangle}.
 \end{aligned}
\end{equation}
%\begin{align}\label{eq:phiIO}
%&\varphi\left(\omega_p\right)=-\mathrm{atan}\Big\{\\
%&\nonumber\left[-2\left(\omega_q-\omega_p\right)g^2m_z-2\left(\omega_0-\omega_p\right)\left(\gamma^2+\left(\omega_q-\omega_p\right)^2\right)\right]/\\
%&\nonumber\left[\kappa\left(\gamma^2+\left(\omega_q-\omega_p\right)^2\right)-\gamma g^2m_z/2\right]\Big\}.
%\end{align}
%where the minus sign in front of the arcus tangens accounts for the phase difference of $\pi$ between the input and the output port ($\lambda/2$ resonator). 
As examples, the phase and amplitude of the bare resonance in Coulomb blockade is simultaneously fit in Fig.~\ref{fig:FigA1}(a) and in Fig.~\ref{fig:Fig_app_shift_linewidth} the same is done for a linecut of Fig.~\ref{fig:Fig4}(a) at $0.25$~T.}
\subsection{Estimation of the photon number}
\label{app:Nphoton}
\corr{Similarly, we may obtain $\langle \hat{a}\rangle(t)$ by solving Eqs.~\eqref{eq:ainoutgen}. Using $\langle \hat{b}_{{\rm in},1}\rangle (t)=\exp(-i\omega_p t)\sqrt{P_{\rm in}/ \omega_p}$, where $P_{\rm in}$ denotes the power in the input field, we find
\begin{equation}
    \label{eq:aoft}
    \langle \hat{a}\rangle (t) = -\sqrt{\frac{\kappa_1 P_{\rm in}}{\hbar\omega_p}}e^{-i\omega_pt}A(\omega_p).
\end{equation}
In the low-drive regime we consider here, we estimate the photon number as
\begin{equation}
    \label{eq:phtonnum}
    n = |\langle \hat{a}\rangle |^2 = \frac{\kappa_1 P_{\rm in}}{\hbar\omega_p}|A(\omega_p)|^2,
\end{equation}
where we approximate $\kappa_1\simeq \kappa/2$.}

%\textcolor{red}{TO BE REPLACED BY PATRICK In Fig.~\ref{fig:Fig3review}, the input-power at the device level is estimated as -133\,dBm by subtracting the calibrated attenuation of the cryostat wiring from the input-power at room temperature. Therefore, we estimate the average number of photons $<0.5$~\cite{palacios2010superconducting,weissl2015kerr} which is an upper bound as internal losses of the resonator and expected reflections at the device level (e.g. wire bonds) are ignored.}
\subsection{Effective two-electron Hamiltonian model}
\label{app:Hamiltonian}
We model an effective two-electron Hamiltonian in the presence of spin-orbit interaction and magnetic field.
We write the Hamiltonian in the basis of singlet and triplet states $\left\{\ket{S_{1,1}},\ket{S_{2,0}},\ket{T_{1,1}^{\pm,0}},\ket{T_{2,0}^{\pm,0}}\right\}$, with the subscripts indicating the charge distribution in the DQD. The Hamiltonian reads
\begin{equation}
\mathcal{H}=\mathcal{H}_0^S+\mathcal{H}_0^T+\mathcal{H}_Z+\mathcal{H}_{\mathrm{so}},
\end{equation}
with the spin quantum-number conserving Hamiltonians
\begin{align}\footnotesize
\mathcal{H}_0^S/\hbar=&-\varepsilon\ket{S_{2,0}}\bra{S_{2,0}}+t_c^S\ket{S_{1,1}}\bra{S_{2,0}}+\text{h.c.}\text{,}\\
\mathcal{H}_0^T/\hbar=&(\Delta_{\rm{ST}}-\varepsilon) \sum_{\pm,0}\nonumber\ket{T_{2,0}^{\pm,0}}\bra{T_{2,0}^{\pm,0}}\\
&+t_c^T\sum_{\pm,0}\ket{T_{1,1}^{\pm,0}}\bra{T_{2,0}^{\pm,0}}+\text{h.c.}\nonumber
\end{align}
\normalsize
Here, $t_{c}^{S,T}$ are the tunnel rates between the two singlets, and between the two triplet states respectively, and $\Delta_\mathrm{ST}$ is the single-dot singlet triplet splitting that separates the $T_{2,0}$ states from the $S_{2,0}$ states.
The Zeeman Hamiltonian is given by
\begin{equation}
\footnotesize
\mathcal{H}_Z/\mu_B =B\sum_\pm\left(\pm \frac{g_L+ g_R}{2}\ket{T_{1,1}^\pm}\bra{T_{1,1}^\pm}\pm g_L\ket{T_{2,0}^\pm}\bra{T_{2,0}^\pm}\right),
\end{equation}
%\normalsize
where $g_L$ ($g_R$) is the Landé g-factor of the left (right) dot. Because of the large intrinsic spin-orbit interaction in the NW, we include the spin-orbit Hamiltonian that couples the singlet and triplet states with opposite charge configuration using the spin-orbit tunnel rate $t_{\rm{SO}}$ as
%\footnotesize
\begin{equation}
\mathcal{H}_{\rm{SO}}/\hbar=t_{\rm{SO}}\left(\ket{T^0_{1,1}}\bra{S_{2,0}}+\sum_{\pm}\pm \ket{T_{1,1}^\pm}\bra{S_{2,0}}\right)+\text{h.c.}
\end{equation}
%\appendix
\section*{Bibliography}
\bibliography{ref}

\begin{thebibliography}{53}
\providecommand{\natexlab}[1]{#1}
\providecommand{\url}[1]{\texttt{#1}}
\expandafter\ifx\csname urlstyle\endcsname\relax
  \providecommand{\doi}[1]{doi: #1}\else
  \providecommand{\doi}{doi: \begingroup \urlstyle{rm}\Url}\fi

\bibitem[Hanson et~al.(2007)Hanson, Kouwenhoven, Petta, Tarucha, and
  Vandersypen]{hanson2007spins}
Ronald Hanson, Leo~P Kouwenhoven, Jason~R Petta, Seigo Tarucha, and Lieven~MK
  Vandersypen.
\newblock Spins in few-electron quantum dots.
\newblock \emph{Reviews of modern physics}, 79\penalty0 (4):\penalty0 1217,
  2007.

\bibitem[Zwanenburg et~al.(2013)Zwanenburg, Dzurak, Morello, Simmons,
  Hollenberg, Klimeck, Rogge, Coppersmith, and Eriksson]{zwanenburg2013silicon}
Floris~A Zwanenburg, Andrew~S Dzurak, Andrea Morello, Michelle~Y Simmons,
  Lloyd~CL Hollenberg, Gerhard Klimeck, Sven Rogge, Susan~N Coppersmith, and
  Mark~A Eriksson.
\newblock Silicon quantum electronics.
\newblock \emph{Reviews of modern physics}, 85\penalty0 (3):\penalty0 961,
  2013.

\bibitem[Vandersypen et~al.(2017)Vandersypen, Bluhm, Clarke, Dzurak, Ishihara,
  Morello, Reilly, Schreiber, and Veldhorst]{vandersypen2017interfacing}
LMK Vandersypen, H~Bluhm, JS~Clarke, AS~Dzurak, R~Ishihara, A~Morello,
  DJ~Reilly, LR~Schreiber, and M~Veldhorst.
\newblock Interfacing spin qubits in quantum dots and donors—hot, dense, and
  coherent.
\newblock \emph{npj Quantum Information}, 3\penalty0 (1):\penalty0 34, 2017.

\bibitem[Chatterjee et~al.(2021)Chatterjee, Stevenson, De~Franceschi, Morello,
  de~Leon, and Kuemmeth]{chatterjee2021semiconductor}
Anasua Chatterjee, Paul Stevenson, Silvano De~Franceschi, Andrea Morello,
  Nathalie~P de~Leon, and Ferdinand Kuemmeth.
\newblock Semiconductor qubits in practice.
\newblock \emph{Nature Reviews Physics}, 3\penalty0 (3):\penalty0 157--177,
  2021.

\bibitem[Childress et~al.(2004)Childress, S{\o}rensen, and
  Lukin]{childress2004mesoscopic}
L~Childress, AS~S{\o}rensen, and Mikhail~D Lukin.
\newblock Mesoscopic cavity quantum electrodynamics with quantum dots.
\newblock \emph{Physical Review A}, 69\penalty0 (4):\penalty0 042302, 2004.

\bibitem[Petersson et~al.(2012)Petersson, McFaul, Schroer, Jung, Taylor, Houck,
  and Petta]{petersson2012circuit}
Karl~D Petersson, Louis~W McFaul, Michael~D Schroer, Minkyung Jung, Jacob~M
  Taylor, Andrew~A Houck, and Jason~R Petta.
\newblock Circuit quantum electrodynamics with a spin qubit.
\newblock \emph{Nature}, 490\penalty0 (7420):\penalty0 380--383, 2012.

\bibitem[Stockklauser et~al.(2017)Stockklauser, Scarlino, Koski, Gasparinetti,
  Andersen, Reichl, Wegscheider, Ihn, Ensslin, and
  Wallraff]{stockklauser2017strong}
Anna Stockklauser, Pasquale Scarlino, Jonne~V Koski, Simone Gasparinetti,
  Christian~Kraglund Andersen, Christian Reichl, Werner Wegscheider, Thomas
  Ihn, Klaus Ensslin, and Andreas Wallraff.
\newblock Strong coupling cavity qed with gate-defined double quantum dots
  enabled by a high impedance resonator.
\newblock \emph{Physical Review X}, 7\penalty0 (1):\penalty0 011030, 2017.

\bibitem[Mi et~al.(2017{\natexlab{a}})Mi, Cady, Zajac, Deelman, and
  Petta]{mi2017strong}
Xiao Mi, JV~Cady, DM~Zajac, PW~Deelman, and Jason~R Petta.
\newblock Strong coupling of a single electron in silicon to a microwave
  photon.
\newblock \emph{Science}, 355\penalty0 (6321):\penalty0 156--158,
  2017{\natexlab{a}}.

\bibitem[Mi et~al.(2018)Mi, Benito, Putz, Zajac, Taylor, Burkard, and
  Petta]{mi2018coherent}
Xiao Mi, M{\'o}nica Benito, Stefan Putz, David~M Zajac, Jacob~M Taylor, Guido
  Burkard, and Jason~R Petta.
\newblock A coherent spin--photon interface in silicon.
\newblock \emph{Nature}, 555\penalty0 (7698):\penalty0 599--603, 2018.

\bibitem[Samkharadze et~al.(2018)Samkharadze, Zheng, Kalhor, Brousse, Sammak,
  Mendes, Blais, Scappucci, and Vandersypen]{samkharadze2018strong}
Nodar Samkharadze, Guoji Zheng, Nima Kalhor, Delphine Brousse, Amir Sammak,
  UC~Mendes, Alexandre Blais, Giordano Scappucci, and LMK Vandersypen.
\newblock Strong spin-photon coupling in silicon.
\newblock \emph{Science}, 359\penalty0 (6380):\penalty0 1123--1127, 2018.

\bibitem[Landig et~al.(2018)Landig, Koski, Scarlino, Mendes, Blais, Reichl,
  Wegscheider, Wallraff, Ensslin, and Ihn]{landig2018coherent}
Andreas~J Landig, Jonne~V Koski, Pasquale Scarlino, UC~Mendes, Alexandre Blais,
  Christian Reichl, Werner Wegscheider, Andreas Wallraff, Klaus Ensslin, and
  T~Ihn.
\newblock Coherent spin--photon coupling using a resonant exchange qubit.
\newblock \emph{Nature}, 560\penalty0 (7717):\penalty0 179--184, 2018.

\bibitem[van Woerkom et~al.(2018)van Woerkom, Scarlino, Ungerer, M{\"u}ller,
  Koski, Landig, Reichl, Wegscheider, Ihn, Ensslin, et~al.]{van2018microwave}
David~J van Woerkom, Pasquale Scarlino, Jann~H Ungerer, Clemens M{\"u}ller,
  Jonne~V Koski, Andreas~J Landig, Christian Reichl, Werner Wegscheider, Thomas
  Ihn, Klaus Ensslin, et~al.
\newblock Microwave photon-mediated interactions between semiconductor qubits.
\newblock \emph{Physical Review X}, 8\penalty0 (4):\penalty0 041018, 2018.

\bibitem[Borjans et~al.(2020)Borjans, Croot, Mi, Gullans, and
  Petta]{borjans2020resonant}
Felix Borjans, XG~Croot, Xiao Mi, MJ~Gullans, and JR~Petta.
\newblock Resonant microwave-mediated interactions between distant electron
  spins.
\newblock \emph{Nature}, 577\penalty0 (7789):\penalty0 195--198, 2020.

\bibitem[Harvey-Collard et~al.(2022)Harvey-Collard, Dijkema, Zheng, Sammak,
  Scappucci, and Vandersypen]{harvey2022coherent}
Patrick Harvey-Collard, Jurgen Dijkema, Guoji Zheng, Amir Sammak, Giordano
  Scappucci, and Lieven~MK Vandersypen.
\newblock Coherent spin-spin coupling mediated by virtual microwave photons.
\newblock \emph{Physical Review X}, 12\penalty0 (2):\penalty0 021026, 2022.

\bibitem[Petta et~al.(2005)Petta, Johnson, Taylor, Laird, Yacoby, Lukin,
  Marcus, Hanson, and Gossard]{petta2005coherent}
Jason~R Petta, Alexander~Comstock Johnson, Jacob~M Taylor, Edward~A Laird, Amir
  Yacoby, Mikhail~D Lukin, Charles~M Marcus, Micah~P Hanson, and Arthur~C
  Gossard.
\newblock Coherent manipulation of coupled electron spins in semiconductor
  quantum dots.
\newblock \emph{Science}, 309\penalty0 (5744):\penalty0 2180--2184, 2005.

\bibitem[Landig et~al.(2019)Landig, Koski, Scarlino, Reichl, Wegscheider,
  Wallraff, Ensslin, and Ihn]{landig2019microwave}
Andreas~J Landig, Jonne~V Koski, Pasquale Scarlino, Christian Reichl, Werner
  Wegscheider, Andreas Wallraff, Klaus Ensslin, and T~Ihn.
\newblock Microwave-cavity-detected spin blockade in a few-electron double
  quantum dot.
\newblock \emph{Physical Review Letters}, 122\penalty0 (21):\penalty0 213601,
  2019.

\bibitem[B{\o}ttcher et~al.(2022)B{\o}ttcher, Harvey, Fallahi, Gardner, Manfra,
  Vool, Bartlett, and Yacoby]{bottcher2022parametric}
CGL B{\o}ttcher, SP~Harvey, Saeed Fallahi, GC~Gardner, MJ~Manfra, Uri Vool,
  SD~Bartlett, and Amir Yacoby.
\newblock Parametric longitudinal coupling between a high-impedance
  superconducting resonator and a semiconductor quantum dot singlet-triplet
  spin qubit.
\newblock \emph{Nature Communications}, 13\penalty0 (1):\penalty0 4773, 2022.

\bibitem[Lehmann et~al.(2013)Lehmann, Wallentin, Jacobsson, Deppert, and
  Dick]{lehmann2013general}
Sebastian Lehmann, Jesper Wallentin, Daniel Jacobsson, Knut Deppert, and
  Kimberly~A Dick.
\newblock A general approach for sharp crystal phase switching in inas, gaas,
  inp, and gap nanowires using only group v flow.
\newblock \emph{Nano letters}, 13\penalty0 (9):\penalty0 4099--4105, 2013.

\bibitem[Nilsson et~al.(2018)Nilsson, Bostr{\"o}m, Lehmann, Dick, Leijnse, and
  Thelander]{nilsson2018tuning}
Malin Nilsson, Florinda~Vi{\~n}as Bostr{\"o}m, Sebastian Lehmann, Kimberly~A
  Dick, Martin Leijnse, and Claes Thelander.
\newblock Tuning the two-electron hybridization and spin states in
  parallel-coupled inas quantum dots.
\newblock \emph{Physical Review Letters}, 121\penalty0 (15):\penalty0 156802,
  2018.

\bibitem[Samkharadze et~al.(2016)Samkharadze, Bruno, Scarlino, Zheng,
  DiVincenzo, DiCarlo, and Vandersypen]{samkharadze2016high}
Nodar Samkharadze, A~Bruno, Pasquale Scarlino, G~Zheng, DP~DiVincenzo,
  L~DiCarlo, and LMK Vandersypen.
\newblock High-kinetic-inductance superconducting nanowire resonators for
  circuit qed in a magnetic field.
\newblock \emph{Physical Review Applied}, 5\penalty0 (4):\penalty0 044004,
  2016.

\bibitem[Ungerer et~al.(2023)Ungerer, Sarmah, Kononov, Ridderbos, Haller,
  Cheung, and Sch{\"o}nenberger]{ungerer2023performance}
Jann~H Ungerer, Deepankar Sarmah, Artem Kononov, Joost Ridderbos, Roy Haller,
  Luk~Yi Cheung, and Christian Sch{\"o}nenberger.
\newblock Performance of high impedance resonators in dirty dielectric
  environments.
\newblock \emph{arXiv preprint arXiv:2302.06303}, 2023.

\bibitem[Borjans et~al.(2021)Borjans, Zhang, Mi, Cheng, Yao, Jackson, Edge, and
  Petta]{borjans2021probing}
Felix Borjans, Xuanzi Zhang, Xiao Mi, Guangming Cheng, Nan Yao, CAC Jackson,
  LF~Edge, and JR~Petta.
\newblock Probing the variation of the intervalley tunnel coupling in a silicon
  triple quantum dot.
\newblock \emph{PRX Quantum}, 2\penalty0 (2):\penalty0 020309, 2021.

\bibitem[Mi et~al.(2017{\natexlab{b}})Mi, P{\'e}terfalvi, Burkard, and
  Petta]{mi2017high}
Xiao Mi, Csaba~G P{\'e}terfalvi, Guido Burkard, and Jason~R Petta.
\newblock High-resolution valley spectroscopy of si quantum dots.
\newblock \emph{Physical review letters}, 119\penalty0 (17):\penalty0 176803,
  2017{\natexlab{b}}.

\bibitem[Ibberson et~al.(2021)Ibberson, Lundberg, Haigh, Hutin, Bertrand,
  Barraud, Lee, Stelmashenko, Oakes, Cochrane, et~al.]{ibberson2021large}
David~J Ibberson, Theodor Lundberg, James~A Haigh, Louis Hutin, Benoit
  Bertrand, Sylvain Barraud, Chang-Min Lee, Nadia~A Stelmashenko, Giovanni~A
  Oakes, Laurence Cochrane, et~al.
\newblock Large dispersive interaction between a cmos double quantum dot and
  microwave photons.
\newblock \emph{PRX Quantum}, 2\penalty0 (2):\penalty0 020315, 2021.

\bibitem[Nilsson et~al.(2016)Nilsson, Namazi, Lehmann, Leijnse, Dick, and
  Thelander]{nilsson2016single}
Malin Nilsson, Luna Namazi, Sebastian Lehmann, Martin Leijnse, Kimberly~A Dick,
  and Claes Thelander.
\newblock Single-electron transport in inas nanowire quantum dots formed by
  crystal phase engineering.
\newblock \emph{Physical Review B}, 93\penalty0 (19):\penalty0 195422, 2016.

\bibitem[{van der Wiel} et~al.(2002){van der Wiel}, De~Franceschi, Elzerman,
  Fujisawa, Tarucha, and Kouwenhoven]{vanderwiel2002}
W.~G. {van der Wiel}, S.~De~Franceschi, J.~M. Elzerman, T.~Fujisawa,
  S.~Tarucha, and L.~P. Kouwenhoven.
\newblock Electron transport through double quantum dots.
\newblock \emph{Reviews of Modern Physics}, 75\penalty0 (1):\penalty0 1--22,
  December 2002.
\newblock \doi{10.1103/RevModPhys.75.1}.

\bibitem[Van~der Wiel et~al.(2002)Van~der Wiel, De~Franceschi, Elzerman,
  Fujisawa, Tarucha, and Kouwenhoven]{van2002electron}
Wilfred~G Van~der Wiel, Silvano De~Franceschi, Jeroen~M Elzerman, Toshimasa
  Fujisawa, Seigo Tarucha, and Leo~P Kouwenhoven.
\newblock Electron transport through double quantum dots.
\newblock \emph{Reviews of modern physics}, 75\penalty0 (1):\penalty0 1, 2002.

\bibitem[Scarlino et~al.(2022)Scarlino, Ungerer, van Woerkom, Mancini, Stano,
  M{\"u}ller, Landig, Koski, Reichl, Wegscheider, et~al.]{scarlino2022situ}
Pasquale Scarlino, Jann~H Ungerer, David~J van Woerkom, Marco Mancini, Peter
  Stano, Clemens M{\"u}ller, Andreas~J Landig, Jonne~V Koski, Christian Reichl,
  Werner Wegscheider, et~al.
\newblock In situ tuning of the electric-dipole strength of a double-dot charge
  qubit: Charge-noise protection and ultrastrong coupling.
\newblock \emph{Physical Review X}, 12\penalty0 (3):\penalty0 031004, 2022.

\bibitem[Blais et~al.(2021)Blais, Grimsmo, Girvin, and
  Wallraff]{blais2021circuit}
Alexandre Blais, Arne~L Grimsmo, Steven~M Girvin, and Andreas Wallraff.
\newblock Circuit quantum electrodynamics.
\newblock \emph{Reviews of Modern Physics}, 93\penalty0 (2):\penalty0 025005,
  2021.

\bibitem[Fasth et~al.(2007)Fasth, Fuhrer, Samuelson, Golovach, and
  Loss]{fasth2007direct}
Carina Fasth, Andreas Fuhrer, Lars Samuelson, Vitaly~N Golovach, and Daniel
  Loss.
\newblock Direct measurement of the spin-orbit interaction in a two-electron
  inas nanowire quantum dot.
\newblock \emph{Physical review letters}, 98\penalty0 (26):\penalty0 266801,
  2007.

\bibitem[Nadj-Perge et~al.(2010)Nadj-Perge, Frolov, Van~Tilburg, Danon,
  Nazarov, Algra, Bakkers, and Kouwenhoven]{nadj2010disentangling}
S~Nadj-Perge, SM~Frolov, JWW Van~Tilburg, J~Danon, Yu~V Nazarov, R~Algra, EPAM
  Bakkers, and LP~Kouwenhoven.
\newblock Disentangling the effects of spin-orbit and hyperfine interactions on
  spin blockade.
\newblock \emph{Physical Review B}, 81\penalty0 (20):\penalty0 201305, 2010.

\bibitem[Trif et~al.(2008)Trif, Golovach, and Loss]{trif2008spin}
Mircea Trif, Vitaly~N Golovach, and Daniel Loss.
\newblock Spin dynamics in inas nanowire quantum dots coupled to a transmission
  line.
\newblock \emph{Physical Review B}, 77\penalty0 (4):\penalty0 045434, 2008.

\bibitem[J{\"u}nger et~al.(2020)J{\"u}nger, Delagrange, Chevallier, Lehmann,
  Dick, Thelander, Klinovaja, Loss, Baumgartner, and
  Sch{\"o}nenberger]{junger2020magnetic}
Christian J{\"u}nger, Rapha{\"e}lle Delagrange, Denis Chevallier, Sebastian
  Lehmann, Kimberly~A Dick, Claes Thelander, Jelena Klinovaja, Daniel Loss,
  Andreas Baumgartner, and Christian Sch{\"o}nenberger.
\newblock Magnetic-field-independent subgap states in hybrid rashba nanowires.
\newblock \emph{Physical Review Letters}, 125\penalty0 (1):\penalty0 017701,
  2020.

\bibitem[Mutter and Burkard(2021)]{mutter2021all}
Philipp~M Mutter and Guido Burkard.
\newblock All-electrical control of hole singlet-triplet spin qubits at
  low-leakage points.
\newblock \emph{Physical Review B}, 104\penalty0 (19):\penalty0 195421, 2021.

\bibitem[Jirovec et~al.(2021)Jirovec, Hofmann, Ballabio, Mutter, Tavani,
  Botifoll, Crippa, Kukucka, Sagi, Martins, et~al.]{jirovec2021singlet}
Daniel Jirovec, Andrea Hofmann, Andrea Ballabio, Philipp~M Mutter, Giulio
  Tavani, Marc Botifoll, Alessandro Crippa, Josip Kukucka, Oliver Sagi,
  Frederico Martins, et~al.
\newblock A singlet-triplet hole spin qubit in planar ge.
\newblock \emph{Nature materials}, 20\penalty0 (8):\penalty0 1106--1112, 2021.

\bibitem[Yu et~al.(2023)Yu, Zihlmann, Abadillo-Uriel, Michal, Rambal,
  Niebojewski, Bedecarrats, Vinet, Dumur, Filippone, et~al.]{yu2023strong}
C{\'e}cile~X Yu, Simon Zihlmann, Jos{\'e}~C Abadillo-Uriel, Vincent~P Michal,
  Nils Rambal, Heimanu Niebojewski, Thomas Bedecarrats, Maud Vinet, {\'E}tienne
  Dumur, Michele Filippone, et~al.
\newblock Strong coupling between a photon and a hole spin in silicon.
\newblock \emph{Nature Nanotechnology}, pages 1--6, 2023.

\bibitem[Burkard et~al.(2021)Burkard, Ladd, Nichol, Pan, and
  Petta]{burkard2021semiconductor}
Guido Burkard, Thaddeus~D Ladd, John~M Nichol, Andrew Pan, and Jason~R Petta.
\newblock Semiconductor spin qubits.
\newblock \emph{arXiv preprint arXiv:2112.08863}, 2021.

\bibitem[Mi et~al.(2017{\natexlab{c}})Mi, Cady, Zajac, Stehlik, Edge, and
  Petta]{mi2017circuit}
X~Mi, JV~Cady, DM~Zajac, J~Stehlik, LF~Edge, and Jason~R Petta.
\newblock Circuit quantum electrodynamics architecture for gate-defined quantum
  dots in silicon.
\newblock \emph{Applied Physics Letters}, 110\penalty0 (4):\penalty0 043502,
  2017{\natexlab{c}}.

\bibitem[Harvey-Collard et~al.(2020)Harvey-Collard, Zheng, Dijkema,
  Samkharadze, Sammak, Scappucci, and Vandersypen]{harvey2020chip}
Patrick Harvey-Collard, Guoji Zheng, Jurgen Dijkema, Nodar Samkharadze, Amir
  Sammak, Giordano Scappucci, and Lieven~MK Vandersypen.
\newblock On-chip microwave filters for high-impedance resonators with
  gate-defined quantum dots.
\newblock \emph{Physical Review Applied}, 14\penalty0 (3):\penalty0 034025,
  2020.

\bibitem[Nilsson et~al.(2017)Nilsson, Chen, Lehmann, Maulerova, Dick, and
  Thelander]{nilsson2017parallel}
Malin Nilsson, I-Ju Chen, Sebastian Lehmann, Vendula Maulerova, Kimberly~A
  Dick, and Claes Thelander.
\newblock Parallel-coupled quantum dots in inas nanowires.
\newblock \emph{Nano letters}, 17\penalty0 (12):\penalty0 7847--7852, 2017.

\bibitem[Ram et~al.(2021)Ram, Persson, Irish, J{\"o}nsson, Timm, and
  Wernersson]{ram2021high}
Mamidala~Saketh Ram, Karl-Magnus Persson, Austin Irish, Adam J{\"o}nsson,
  Rainer Timm, and Lars-Erik Wernersson.
\newblock High-density logic-in-memory devices using vertical indium arsenide
  nanowires on silicon.
\newblock \emph{Nature Electronics}, 4\penalty0 (12):\penalty0 914--920, 2021.

\bibitem[Forn-D{\'\i}az et~al.(2019)Forn-D{\'\i}az, Lamata, Rico, Kono, and
  Solano]{forn2019ultrastrong}
P~Forn-D{\'\i}az, L~Lamata, E~Rico, J~Kono, and E~Solano.
\newblock Ultrastrong coupling regimes of light-matter interaction.
\newblock \emph{Reviews of Modern Physics}, 91\penalty0 (2):\penalty0 025005,
  2019.

\bibitem[Frey et~al.(2012)Frey, Leek, Beck, Blais, Ihn, Ensslin, and
  Wallraff]{frey2012dipole}
T~Frey, PJ~Leek, M~Beck, Alexandre Blais, Thomas Ihn, Klaus Ensslin, and
  Andreas Wallraff.
\newblock Dipole coupling of a double quantum dot to a microwave resonator.
\newblock \emph{Physical Review Letters}, 108\penalty0 (4):\penalty0 046807,
  2012.

\bibitem[Crippa et~al.(2019)Crippa, Ezzouch, Apr{\'a}, Amisse, Lavi{\'e}ville,
  Hutin, Bertrand, Vinet, Urdampilleta, Meunier, et~al.]{crippa2019gate}
Alessandro Crippa, R~Ezzouch, A~Apr{\'a}, A~Amisse, R~Lavi{\'e}ville, L~Hutin,
  B~Bertrand, M~Vinet, Matias Urdampilleta, Tristan Meunier, et~al.
\newblock Gate-reflectometry dispersive readout and coherent control of a spin
  qubit in silicon.
\newblock \emph{Nature communications}, 10\penalty0 (1):\penalty0 1--6, 2019.

\bibitem[Ezzouch et~al.(2021)Ezzouch, Zihlmann, Michal, Li, Apr{\'a}, Bertrand,
  Hutin, Vinet, Urdampilleta, Meunier, et~al.]{ezzouch2021dispersively}
Rami Ezzouch, Simon Zihlmann, Vincent~P Michal, Jing Li, Agostino Apr{\'a},
  Benoit Bertrand, Louis Hutin, Maud Vinet, Matias Urdampilleta, Tristan
  Meunier, et~al.
\newblock Dispersively probed microwave spectroscopy of a silicon hole double
  quantum dot.
\newblock \emph{Physical Review Applied}, 16\penalty0 (3):\penalty0 034031,
  2021.

\bibitem[Park et~al.(2020)Park, Metzger, Tosi, Goffman, Urbina, Pothier, and
  Yeyati]{park2020adiabatic}
Sunghun Park, Cyril Metzger, Leandro Tosi, Marcelo~Fabi{\'a}n Goffman,
  C~Urbina, Hugues Pothier, and A~Levy Yeyati.
\newblock From adiabatic to dispersive readout of quantum circuits.
\newblock \emph{Physical Review Letters}, 125\penalty0 (7):\penalty0 077701,
  2020.

\bibitem[Schroer et~al.(2012)Schroer, Jung, Petersson, and
  Petta]{schroer2012radio}
MD~Schroer, M~Jung, KD~Petersson, and Jason~R Petta.
\newblock Radio frequency charge parity meter.
\newblock \emph{Physical review letters}, 109\penalty0 (16):\penalty0 166804,
  2012.

\bibitem[Malinowski et~al.(2018)Malinowski, Martins, Smith, Bartlett, Doherty,
  Nissen, Fallahi, Gardner, Manfra, Marcus, et~al.]{malinowski2018spin}
Filip~K Malinowski, Frederico Martins, Thomas~B Smith, Stephen~D Bartlett,
  Andrew~C Doherty, Peter~D Nissen, Saeed Fallahi, Geoffrey~C Gardner,
  Michael~J Manfra, Charles~M Marcus, et~al.
\newblock Spin of a multielectron quantum dot and its interaction with a
  neighboring electron.
\newblock \emph{Physical Review X}, 8\penalty0 (1):\penalty0 011045, 2018.

\bibitem[Shore and Knight(1993)]{shore1993jaynes}
Bruce~W Shore and Peter~L Knight.
\newblock The jaynes-cummings model.
\newblock \emph{Journal of Modern Optics}, 40\penalty0 (7):\penalty0
  1195--1238, 1993.

\bibitem[Blais et~al.(2004)Blais, Huang, Wallraff, Girvin, and
  Schoelkopf]{blais2004cavity}
Alexandre Blais, Ren-Shou Huang, Andreas Wallraff, Steven~M Girvin, and R~Jun
  Schoelkopf.
\newblock Cavity quantum electrodynamics for superconducting electrical
  circuits: An architecture for quantum computation.
\newblock \emph{Physical Review A}, 69\penalty0 (6):\penalty0 062320, 2004.

\bibitem[Gardiner and Zoller(2004)]{gardiner2004quantum}
Crispin Gardiner and Peter Zoller.
\newblock \emph{Quantum noise: a handbook of Markovian and non-Markovian
  quantum stochastic methods with applications to quantum optics}.
\newblock Springer Science \& Business Media, 2004.

\bibitem[Wong and Vavilov(2017)]{wong:2017}
C.~H. Wong and M.~G. Vavilov.
\newblock Quantum efficiency of a single microwave photon detector based on a
  semiconductor double quantum dot.
\newblock \emph{Phys. Rev. A}, 95:\penalty0 012325, Jan 2017.
\newblock \doi{10.1103/PhysRevA.95.012325}.

\bibitem[Schöndorf et~al.(2018)Schöndorf, Govia, Vavilov, McDermott, and
  Wilhelm]{schondorf:2018}
M.~Schöndorf, L.~C.~G. Govia, M.~G. Vavilov, R.~McDermott, and F.~K. Wilhelm.
\newblock Optimizing microwave photodetection: input–output theory.
\newblock \emph{Quantum Sci. Technol.}, 3\penalty0 (2):\penalty0 024009, mar
  2018.
\newblock \doi{10.1088/2058-9565/aaa7f7}.

\end{thebibliography}


\begin{thebibliography}{13}
\providecommand{\natexlab}[1]{#1}
\providecommand{\url}[1]{\texttt{#1}}
\expandafter\ifx\csname urlstyle\endcsname\relax
  \providecommand{\doi}[1]{doi: #1}\else
  \providecommand{\doi}{doi: \begingroup \urlstyle{rm}\Url}\fi

\bibitem[Barker et~al.(2019)Barker, Lehmann, Namazi, Nilsson, Thelander, Dick,
  and Maisi]{barker2019individually}
David Barker, Sebastian Lehmann, Luna Namazi, Malin Nilsson, Claes Thelander,
  Kimberly~A Dick, and Ville~F Maisi.
\newblock Individually addressable double quantum dots formed with nanowire
  polytypes and identified by epitaxial markers.
\newblock \emph{Applied Physics Letters}, 114\penalty0 (18), 2019.

\bibitem[Gatzke et~al.()Gatzke, Webb, Fobelets, and Stradling]{gatzke1998}
C.~Gatzke, S.~J. Webb, K.~Fobelets, and R.~A. Stradling.
\newblock In situ {{Raman}} spectroscopy of the selective etching of
  antimonides in {{GaSb}}/{{AlSb}}/{{InAs}} heterostructures.
\newblock 13\penalty0 (4):\penalty0 399.
\newblock ISSN 0268-1242.
\newblock \doi{10.1088/0268-1242/13/4/008}.
\newblock URL \url{https://dx.doi.org/10.1088/0268-1242/13/4/008}.

\bibitem[Nilsson et~al.(2017)Nilsson, Chen, Lehmann, Maulerova, Dick, and
  Thelander]{nilsson2017parallel}
Malin Nilsson, I-Ju Chen, Sebastian Lehmann, Vendula Maulerova, Kimberly~A
  Dick, and Claes Thelander.
\newblock Parallel-coupled quantum dots in inas nanowires.
\newblock \emph{Nano letters}, 17\penalty0 (12):\penalty0 7847--7852, 2017.

\bibitem[Thelander et~al.(2011)Thelander, Caroff, Plissard, Dey, and
  Dick]{thelander2011effects}
Claes Thelander, Philippe Caroff, S{\'e}bastien Plissard, Anil~W Dey, and
  Kimberly~A Dick.
\newblock Effects of crystal phase mixing on the electrical properties of inas
  nanowires.
\newblock \emph{Nano letters}, 11\penalty0 (6):\penalty0 2424--2429, 2011.

\bibitem[{van der Wiel} et~al.(2002){van der Wiel}, De~Franceschi, Elzerman,
  Fujisawa, Tarucha, and Kouwenhoven]{vanderwiel2002}
W.~G. {van der Wiel}, S.~De~Franceschi, J.~M. Elzerman, T.~Fujisawa,
  S.~Tarucha, and L.~P. Kouwenhoven.
\newblock Electron transport through double quantum dots.
\newblock \emph{Reviews of Modern Physics}, 75\penalty0 (1):\penalty0 1--22,
  December 2002.
\newblock \doi{10.1103/RevModPhys.75.1}.

\bibitem[Van~der Wiel et~al.(2002)Van~der Wiel, De~Franceschi, Elzerman,
  Fujisawa, Tarucha, and Kouwenhoven]{van2002electron}
Wilfred~G Van~der Wiel, Silvano De~Franceschi, Jeroen~M Elzerman, Toshimasa
  Fujisawa, Seigo Tarucha, and Leo~P Kouwenhoven.
\newblock Electron transport through double quantum dots.
\newblock \emph{Reviews of modern physics}, 75\penalty0 (1):\penalty0 1, 2002.

\bibitem[Scarlino et~al.(2022)Scarlino, Ungerer, van Woerkom, Mancini, Stano,
  M{\"u}ller, Landig, Koski, Reichl, Wegscheider, et~al.]{scarlino2022situ}
Pasquale Scarlino, Jann~H Ungerer, David~J van Woerkom, Marco Mancini, Peter
  Stano, Clemens M{\"u}ller, Andreas~J Landig, Jonne~V Koski, Christian Reichl,
  Werner Wegscheider, et~al.
\newblock In situ tuning of the electric-dipole strength of a double-dot charge
  qubit: Charge-noise protection and ultrastrong coupling.
\newblock \emph{Physical Review X}, 12\penalty0 (3):\penalty0 031004, 2022.

\bibitem[Blais et~al.(2004)Blais, Huang, Wallraff, Girvin, and
  Schoelkopf]{blais2004cavity}
Alexandre Blais, Ren-Shou Huang, Andreas Wallraff, Steven~M Girvin, and R~Jun
  Schoelkopf.
\newblock Cavity quantum electrodynamics for superconducting electrical
  circuits: An architecture for quantum computation.
\newblock \emph{Physical Review A}, 69\penalty0 (6):\penalty0 062320, 2004.

\bibitem[Fasth et~al.(2007)Fasth, Fuhrer, Samuelson, Golovach, and
  Loss]{fasth2007direct}
Carina Fasth, Andreas Fuhrer, Lars Samuelson, Vitaly~N Golovach, and Daniel
  Loss.
\newblock Direct measurement of the spin-orbit interaction in a two-electron
  inas nanowire quantum dot.
\newblock \emph{Physical review letters}, 98\penalty0 (26):\penalty0 266801,
  2007.

\bibitem[Nadj-Perge et~al.(2010)Nadj-Perge, Frolov, Van~Tilburg, Danon,
  Nazarov, Algra, Bakkers, and Kouwenhoven]{nadj2010disentangling}
S~Nadj-Perge, SM~Frolov, JWW Van~Tilburg, J~Danon, Yu~V Nazarov, R~Algra, EPAM
  Bakkers, and LP~Kouwenhoven.
\newblock Disentangling the effects of spin-orbit and hyperfine interactions on
  spin blockade.
\newblock \emph{Physical Review B}, 81\penalty0 (20):\penalty0 201305, 2010.

\bibitem[Nilsson et~al.(2018)Nilsson, Bostr{\"o}m, Lehmann, Dick, Leijnse, and
  Thelander]{nilsson2018tuning}
Malin Nilsson, Florinda~Vi{\~n}as Bostr{\"o}m, Sebastian Lehmann, Kimberly~A
  Dick, Martin Leijnse, and Claes Thelander.
\newblock Tuning the two-electron hybridization and spin states in
  parallel-coupled inas quantum dots.
\newblock \emph{Physical Review Letters}, 121\penalty0 (15):\penalty0 156802,
  2018.

\bibitem[Trif et~al.(2008)Trif, Golovach, and Loss]{trif2008spin}
Mircea Trif, Vitaly~N Golovach, and Daniel Loss.
\newblock Spin dynamics in inas nanowire quantum dots coupled to a transmission
  line.
\newblock \emph{Physical Review B}, 77\penalty0 (4):\penalty0 045434, 2008.

\bibitem[J{\"u}nger et~al.(2020)J{\"u}nger, Delagrange, Chevallier, Lehmann,
  Dick, Thelander, Klinovaja, Loss, Baumgartner, and
  Sch{\"o}nenberger]{junger2020magnetic}
Christian J{\"u}nger, Rapha{\"e}lle Delagrange, Denis Chevallier, Sebastian
  Lehmann, Kimberly~A Dick, Claes Thelander, Jelena Klinovaja, Daniel Loss,
  Andreas Baumgartner, and Christian Sch{\"o}nenberger.
\newblock Magnetic-field-independent subgap states in hybrid rashba nanowires.
\newblock \emph{Physical Review Letters}, 125\penalty0 (1):\penalty0 017701,
  2020.

\end{thebibliography}
\clearpage
\section{Extended data}
\clearpage
\end{document}

\preprint{APS/123-QED}
\bibliographystyle{unsrtnat} 
\newcommand{\corr}[1]{\textcolor{black}{#1}}
%=======================================================================================================================================
\title{Supplementary to 'Strong coupling between a microwave photon and a singlet-triplet qubit'}
%%%%%
\author{J.\,H.~Ungerer}
\altaffiliation{Equal contributions.}
%\email{jannhinnerk.ungerer@unibas.ch}
\affiliation{
Department of Physics, University of Basel, Klingelbergstrasse 82 CH-4056, Switzerland
}
\affiliation{
Swiss Nanoscience Institute, University of Basel, Klingelbergstrasse 82 CH-4056, Switzerland
}
\author{A.~Pally\footnotemark[1]}
\altaffiliation{Equal contributions.}
\affiliation{
Department of Physics, University of Basel, Klingelbergstrasse 82 CH-4056, Switzerland
}
\author{A.~Kononov}
\affiliation{
Department of Physics, University of Basel, Klingelbergstrasse 82 CH-4056, Switzerland
}

\author{S.~Lehmann}
\affiliation{Solid State Physics and NanoLund, Lund University, Box 118, S-22100 Lund, Sweden}

\author{J.~Ridderbos}
\altaffiliation{Current address: MESA Institute for Nanotechnology, University of Twente, P.O. Box 217, 7500 AE Enschede, The
Netherlands}
\affiliation{
Department of Physics, University of Basel, Klingelbergstrasse 82 CH-4056, Switzerland
}

\author{P.\,P. Potts}
\affiliation{
Department of Physics, University of Basel, Klingelbergstrasse 82 CH-4056, Switzerland
}

\author{C.~Thelander}
\affiliation{Solid State Physics and NanoLund, Lund University, Box 118, S-22100 Lund, Sweden}
\author{K.A.~Dick}
\affiliation{Centre for Analysis and Synthesis, Lund University, Box 124, S-22100 Lund, Sweden}
\author{V.F.~Maisi}
\affiliation{Solid State Physics and NanoLund, Lund University, Box 118, S-22100 Lund, Sweden}
\author{P.~Scarlino}
\affiliation{Institute of Physics and Center for Quantum Science and Engineering, Ecole Polytechnique Fédérale de Lausanne, CH-1015 Lausanne, Switzerland}
\author{A.~Baumgartner}
\homepage{www.nanoelectronics.unibas.ch}
\affiliation{
Department of Physics, University of Basel, Klingelbergstrasse 82 CH-4056, Switzerland
}
\affiliation{
Swiss Nanoscience Institute, University of Basel, Klingelbergstrasse 82 CH-4056, Switzerland
}
\author{C.~Sch{\"o}nenberger}
\homepage{www.nanoelectronics.unibas.ch}
\affiliation{
Department of Physics, University of Basel, Klingelbergstrasse 82 CH-4056, Switzerland
}
\affiliation{
Swiss Nanoscience Institute, University of Basel, Klingelbergstrasse 82 CH-4056, Switzerland
}
\date{\today}

%%% To add S in front of figures
\renewcommand{\thefigure}{S\arabic{figure}}
\setcounter{figure}{0}

\maketitle
\section{InAs crystal-phase nanowires}
InAs nanowires with controlled crystal structure were grown by metal-organic vapor phase epitaxy (MOVPE) from Au aerosol nanoparticles with a nominal diameter of 30 nm deposited on InAs 111B substrates. After annealing, nanowires were grown at 460\,°C by introducing trimethylindium (TMIn) at a molar fraction of $\chi_\mathrm{TMIn} = 1.8\cdot 10^{-6}$ and Arsine (AsH\textsubscript{3}) at a molar fraction of $\chi_\mathrm{AsH\textsubscript{3}}= 1.2\cdot 10^{-4}$. Crystal-phase switching is realized by modifying the AsH\textsubscript{3} molar fractions from $\chi_\mathrm{AsH\textsubscript{3}} = 2.5\cdot 10^{-2}$ for zinc blende to $\chi_\mathrm{AsH\textsubscript{3}} = 2.2\cdot 10^{-5}$ for wurtzite, with 15\,s waiting steps under AsH\textsubscript{3}. The wurtzite barrier growth time is 54\,s and the zinc blende segment growth time is 360\,s. Deposited at the same growth temperature, the GaSb shell was grown for 40 minutes with respective molar fractions of trimethylgallium (TMGa) $\chi_\mathrm{TMGa} = 2.7\cdot 10^{-6}$ and trimethylantimony (TMSb) $\chi_\mathrm{TMSb}= 3.1\cdot 10^{-5}$. 
As the nanowires in this work were grown from randomly deposited Au seed particles, a variability in the local growth conditions was present that affect nanowire growth rate and segment lengths.
However, by adding the GaSb shell that selectively deposits on zinc blende surfaces, it is possible to identify nanowires with desired segment lengths, and to accurately position contacts and local gates to these~\cite{barker2019individually}. The GaSb-shell is then removed before contacting by a wet-etching process using MF-319 developer~\cite{gatzke1998,barker2019individually}.

By growing nanowires in arrays, such as from lithographically defined Au particles, the variability in segment lengths can be greatly reduced~\cite{nilsson2017parallel}. The electron mobility in these nanowires is primarily limited by the wurtzite tunnel barriers and surface scattering. InAs nanowires with a pure zinc-blende crystal phase grown by a corresponding method show a room-temperature field-effect mobility of approximately 2000 cm\textsuperscript{2}$/$Vs~\cite{thelander2011effects}.

In total, we have fabricated 4 nanowire devices coupled to a high-impedance resonator. All nanowires demonstrated well-defined double-quantum dots as expected from their barrier design. Out of these 4 devices, two were investigated at elevated magnetic-field strengths and both of them showed similar behavior as discussed in the manuscript. 
\section{Hamiltonian in the odd charge parity}
In the main text, we elaborate on the Hamiltonian describing the double quantum dot (DQD) for an even charge occupation. 
This section provides the description for an \textit{odd} number of electrons which is used in order to obtain Fig.~6(d) in the extended data.
In this case, the total electron spin is $1/2$ which can be modelled by one electron with a half spin.
This electron can reside either on the left dot or on the right dot~\cite{vanderwiel2002}. Therefore, a suitable basis is $\left\{\ket{\mathrm{L}\uparrow},\ket{\mathrm{L}\downarrow},\ket{\mathrm{R}\uparrow},\ket{\mathrm{R}\downarrow}\right\}$, where L/R denotes whether the charge resides in the  left dot or on the right dot, and $\uparrow$/$\downarrow$ denotes whether the spin is aligned parallel or anti-parallel with the magnetic field $B$.

The Hamiltonian describing the electron can be decomposed into three parts as
\begin{equation}
\mathcal{H}_\mathrm{odd}=\mathcal{H}_\mathrm{odd}^0+\mathcal{H}_\mathrm{odd}^Z+\mathcal{H}_\mathrm{odd}^\mathrm{SO}
\end{equation}
The first part of the Hamiltonian describes the spin-independent charge which can be written using the the charge Pauli matrices $\hat{\tau}_{x,y,z}$ as
\begin{equation}
\mathcal{H}_\mathrm{odd}^0=\frac{\hbar\epsilon}{2}\hat{\tau}_z+\hbar t_c\hat{\tau}_x\text{.}
\end{equation}
Here, the diagonal terms are proportional to the detuning $\hbar\epsilon=E_R-E_L$ which is the difference between the electro-static potential of the electron residing in the right and left dot.
The off-diagonal terms are given by $\hbar t_c$, which is the spin-conserving tunnel rate.
%In the absence of a magnetic-field and spin-orbit interaction, $\mathcal{H}_\mathrm{odd}=H_\mathrm{odd}^0$ is the full Hamiltonian. The resulting energy levels are plotted in Fig.~\ref{fig:levels_odd}a).
%As visible in the figure, at $\epsilon=0$, the spin-degenerate charge states hybridize.
%At this charge degeneracy, the eigenstates of the system are given by the the anti-symmetric (bonding) and symmetric (anti-bonding) superposition states, $\ket{\pm_c}=(\ket{R\updownarrow}\pm\ket{L\updownarrow})/\sqrt{2}$.

In the presence of a magnetic-field, $\mathcal{H}_\mathrm{odd}^Z$ comes into effect.
This term describes the Zeeman energy of the electron and is given by
\begin{equation}
\mathcal{H}_\mathrm{odd}^Z=\frac{1}{2}g_{L,R}\mu_B B\hat{\sigma}_z\text{,}
\end{equation}
where $g_L$ and $g_R$ are the site-dependent Landé g-factors, $\mu_B$ is the Bohr magneton and $\hat{\sigma}_{x,y,z}$ are the spin Pauli matrices.
The Zeeman energy lifts the spin degeneracy and hence four spin-polarized levels are observed as shown in Fig.~6(d) in the extended data.
As explained in the methods section, unequal g factors $g_L \neq g_R$ result in a shift of the avoided level crossings originating from spin-conserving tunneling.
This results in a slope of the observed inter-dot transition as a function of gate voltage (detuning) and field from zero field onward.
\section{Analysis of device B}
\begin{figure}
    \centering
    \includegraphics[width=\linewidth]{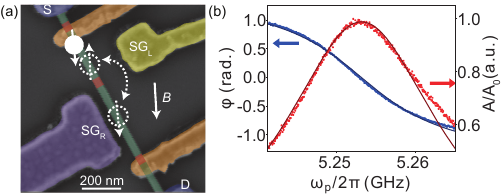}
    \caption{\textbf{Device B}
    (a) False colored SEM-image of the device. The NW (green) is divided into two segments by an in-situ grown tunnel barrier (red), thus forming the DQD system. The NW ends are contacted by two Ti/Au contacts (S,D) and the NW segements can be electrically tuned by two Ti/Au sidegates SG$_R$ (purple) and SG$_L$ (yellow).
    A high-impedance, half-wave resonator is connected to SG$_R$. Top gates (orange) are kept at a constant voltage of $-0.05$~V. The magnetic field is applied in-plane at an angle of $\sim60^\circ$ %$123.6^\circ$
    to the NW. The arrows illustrate an even charge configuration with the two degenerate DQD states $T_{1,1}^+$ and $S_{2,0}$.
    (b)  Resonance curve of the resonator in Coulomb blockade in amplitude $A/A_0$ (red) and phase $\varphi$ (blue).}
    \label{fig:deviceB}
\end{figure}
In this section we will discuss device B, which showed qualitatively similar behavior as device A which is discussed in the main text. Device B is shown in Fig.~\ref{fig:deviceB}(a), including a false-colored SEM-image of the crystal-phase defined NW DQD. The DQD is hosted in the $280$\,nm and $380$\,nm long zincblende segments (green), separated by $30$\,nm long wurtzite (red) tunnel barriers. A high-impedance, half-wave coplanar-waveguide resonator is capacitively coupled to the DQD at its voltage anti-node via \corr{sidegate $SG_R$.} One more side gate $SG_L$ (yellow) allows to tune the electrostatic potential and there are two top gates (orange) kept at constant voltage of $-0.05$.
\corr{We  show the bare resonance curve in Fig.~\ref{fig:deviceB}(b) and extract the bare resonance frequency $\omega_0/2\pi=5.25308\pm0.00003$\,GHz and the bare decay rate \corr{$\kappa/2\pi=23.2\pm0.8$} MHz.} 
\corr{The main difference to device A is the weaker coupling of the resonator gate $SG_R$ to the DQD and the smaller voltages applied to the top gates, which resulted in a stronger coupling of the DQD to the leads. Consequently, we observe a weaker spin-photon coupling and a larger qubit linewidth compared to device A. However, the behaviour of the singlet-triplet qubit in magnetic field is qualitatively the same, demonstrating that this kind of singlet-triplet is reproducible.}
\subsection{Charge-stability diagram}
\begin{figure}
    \centering    \includegraphics[width=\linewidth]{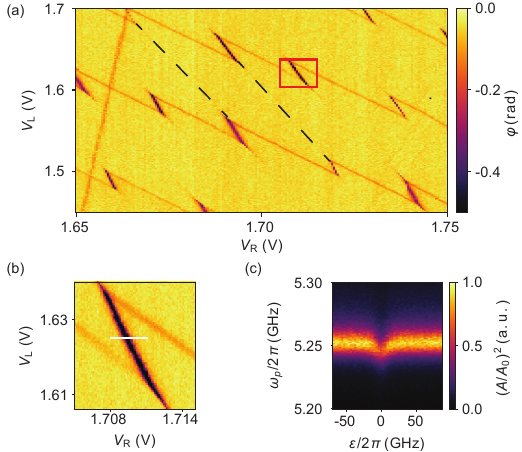}\caption{\textbf{Dispersive sensing of the DQD at $\mathbf{B=0}$.}
    (a) Charge stability diagram of the device, in which the resonator phase $\varphi$ is measured as a function of the SG voltages $V_R$ and $V_L$.
    The negative slopes of the interdot transitions are due to the strong cross-capacitance of the larger gate $SG_R$. A zoom on the interdot transition pointed out by the red rectangle is shown in (b). (c) Resonator transmission $(A/A_0)^2$ versus probe frequency $
    \omega_\mathrm{p}$ and detuning $\epsilon$ along the white line in (b). At the charge degeneracy point of the DQD, we observe a dispersive shift with respect to the bare resonance frequency.}
    \label{fig:Fig2}
\end{figure}
Fig.~\ref{fig:Fig2}(a) shows the charge stability diagram of the DQD detected as a shift in the transmission phase $\varphi$ of the resonator, plotted as a function of the two gate voltages $V_L$ and $V_R$ at a fixed probe frequency of $5.253$ GHz, close to resonance.
%i.e detuned from the bare resonace frequency by $80$ kHz.
We observe a slanted honeycomb pattern, in which the inter-dot transition lines exhibit a negative slope due to the specific gate geometry (see Fig.~\ref{fig:deviceB}(a)), which results in the right gate ($V_R$) coupling stronger to both dots than the left ($V_L$).
Using a capacitance model~\cite{van2002electron,scarlino2022situ}, we extract the gate-to-dot capacitances $C_{R2} = 2.5\pm0.4$\,aF, $C_{L2} =1.65\pm0.08$\,aF, $C_{R1}=10.1\pm0.6 $\,aF and $C_{L1} =2.0\pm0.2$\,aF.

In Fig.~\ref{fig:Fig2}(c) we show the resonator response while varying the probe frequency $\omega_p$ and changing the detuning $\epsilon$ along the white line in Fig.~\ref{fig:Fig2}(b).
By fitting \corr{input-output theory} to this particular IDT, we extract the inter-dot tunnel coupling $t = 4.40\pm0.06$ GHz, charge-photon coupling $g_0 = 150\pm3$ MHz, and charge qubit linewidth $\gamma = 1.5\pm0.5$~GHz.
\subsection{Avoided crossing}
\label{sec:anti-crossing}
\begin{figure}
\centering
\includegraphics{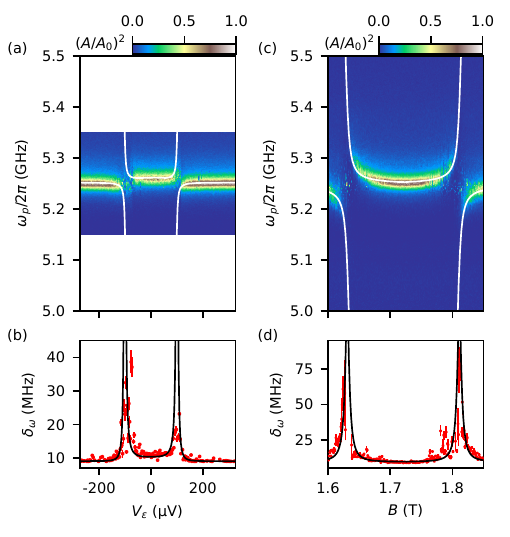}\caption{\textbf{Avoided crossing as function of gate voltage and field.} (a) Resonator transmission at a fixed magnetic field $B=1.67$\,T as function of detuning gate voltage $V_\epsilon$. (b) Linewidths $\delta_\omega$ extracted from (a). (c) Resonator transmission at a fixed detuning voltage corresponding to $\epsilon^\prime=1.65$\,GHz as function of magnetic field. (d) Linewidths $\delta_\omega$ extracted from (c). Using a Jaynes-Cummings model fit, we extract the following parameters for (a) and (b): $g_0/2\pi=164\pm6$\,MHz, $t/2\pi=1.34\pm0.05$\,GHz, $\gamma=317\pm28$\,MHz, $\kappa=18.2\pm0.2$\,MHz. This results in a resonant coupling strength of $g(\omega_q=\omega_0)=83\pm4$\,MHz when correcting for the mixing angle. From (c) and (d), we extract $g(\omega_q=\omega_0)/2\pi=158\pm3$\,MHz, $t(B=B_0)/2\pi\sim 0$, $\gamma/2\pi=269\pm16$\,MHz, $\kappa/2\pi=18.6\pm0.2$\,MHz. The larger value of the coupling strength in the magnetic-field (c,d) sweep compared to the detuning sweep (a,b) is attributed to the smaller mixing angle and reflected by the larger splitting at the anti-crossing in (c) compared to (a). Given the input power $P_\mathrm{in}=-133$\,dBm, the average number of photons in these measurements is $n<0.25$ (see methods of main part). \label{fig:Fig4}\label{fig:Fig3review}}
\end{figure}
As illustrated in Fig.~\ref{fig:Fig3}(c), the DQD can be operated as a singlet-triplet qubit when placed into a magnetic field. The qubit frequency $\omega_q=\Delta_{SO}/\hbar$ can be brought into resonance with the cavity frequency $\omega_0$ at $B\approx 1.7$~T, as discussed in more detail below. At the resonance condition ($\omega_q\sim\omega_0$), an anti-symmetric (bonding) and a symmetric (anti-bonding) qubit-photon superposition are formed.
Consistently, at a field of $B \approx 1.7$\,T, we observe an anti-crossing between the resonator and the singlet-triplet qubit.
Figure~\ref{fig:Fig4}(a) shows the anti-crossing as a function of the detuning voltage $V_\epsilon$ at constant magnetic field $B=1.67$\,T. By fitting a lorentzian to each trace of fixed detuning, we extract the resonance frequencies $\omega_{\Psi_\pm}$ and linewidths $\delta_\omega$ (\corr{transmission} and phase).
Simultaneously, we fit the transition frequencies (dashed, white curves in Fig.~\ref{fig:Fig4}(a)) and linewidths (solid, black curve in Fig.~\ref{fig:Fig4}(b)) to the Jaynes-Cummings model. \corr{The transition frequencies are fitted as described in the methods section in the main text and linewidth of the transitions from the ground state to the predominantly photon-like dressed state $\ket{\psi_-}$ is given by}
\begin{align}
\delta_\omega&=\left|\langle \psi_- |g,1\rangle \right|^2\kappa+\left|\langle\psi_-|e,0\rangle\right|^22\gamma\label{eq:JClinewidth}\\
&=\cos^2\left(\theta\right)\kappa+\sin^2 \left(\theta\right) 2\gamma,\nonumber
\end{align}
where $\theta=\frac{1}{2}\tan^{-1}\left(\frac{2g}{\omega_q-\omega_0 }\right)$~\cite{blais2004cavity}.
The fit parameters are given in the caption of Fig.~\ref{fig:Fig3review}.

%We note that the observed anti-crossing occurs at a finite detuning $\epsilon^\prime\sim4.8$\,GHz and hence DQD polarization, which reduces the resonator-qubit coupling strength as $g(\epsilon)=g_0\sin\theta$, with $\sin \theta=2t/\sqrt{(2t)^2+{\epsilon}^2}$, where $\theta$ is the mixing angle~\cite{childress2004mesoscopic,petersson2012circuit,frey2012dipole}.
%After correcting for this mixing angle we obtain a spin-photon coupling strength $g(\epsilon=\epsilon^\prime)/2\pi=83\pm4\rm{MHz}$.}
In Fig.~\ref{fig:Fig4}(c), we show the same anti-crossing as a function of $B$ at a fixed detuning $\epsilon/2\pi\sim 1.65$\,GHz.
To extract the spin-photon coupling strength and qubit linewidth from this second measurement, we characterize the effective qubit transition frequency around the minimum $t_0=t(B_0)$ by $\omega_q(B)=\sqrt{(2t_0)^2+(\alpha_B(B-B_0))^2}$, where we introduce the heuristic scaling factor $\alpha_B$ .
With this additional free parameter, we fit the Jaynes-Cummings model (dashed, white curves in Fig. 3(c) and solid, black curve in Fig. 3(d)) and extract the parameters described in the caption of Fig.~\ref{fig:Fig3review}.
%The extracted spin-photon coupling strength on resonance $g(\omega_q=\omega_0)/2\pi = 158 \pm 3$ MHz, qubit linewidth $\gamma/2\pi=269\pm16\rm{\,MHz}$ and resonator decay rate $\kappa/2\pi=18.6\pm0.2$\,MHz for Fig.~\ref{fig:Fig4}(c,d), demonstrate that the strong coupling regime can be reached.
\subsection{Magnetospectroscopy}
\begin{figure}
    \centering
    \includegraphics{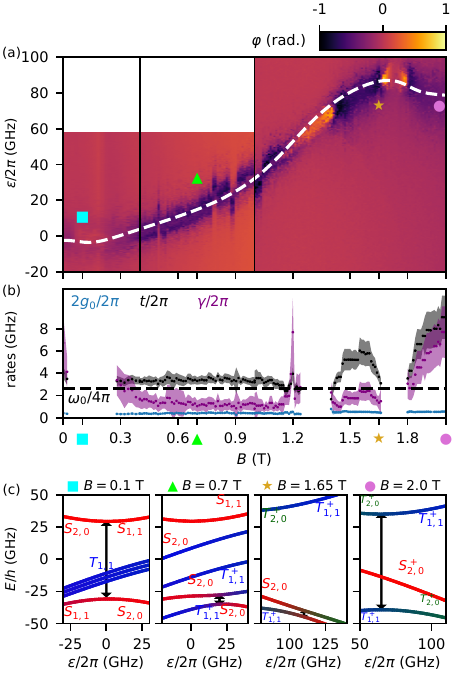}
    \caption{\textbf{Magnetospectroscopy of the singlet-triplet qubit.} a) Resonator \corr{phase $\varphi$} as a function of the magnetic field $B$ and detuning $\epsilon$. The white dashed line is a fit of the effective two-electron Hamiltonian to the data. b) Extracted tunnel rate $2t/2\pi$ (black), twice the qubit-photon coupling $\corr{2g_0}/2\pi$ (\corr{cyan}) and qubit linewidth $\gamma/2\pi$ (\corr{purple}). \corr{Half} the bare resonator frequency is indicated by the \corr{black dashed} line. (c) Two-electron energy level diagrams at various magnetic fields with the corresponding field strength indicated in (a) and (b) by the given symbols. A constant offset of $20$\,GHz and $30$\,GHz was \corr{added to} the energy levels at 1.65\,T and 2.0\,T, respectively.}
    %Given the input-power $P_\mathrm{in}=-118$\,dBm in these experiments, the average photon number is $n\lesssim8$ (see methods in main text).}
    \label{fig:Fig3}
\end{figure}

\corr{In this section we analyse the magnetospectroscopy of device B analogously to device A. We measure the resonator phase $\varphi$ as a function of the detuning $\epsilon$ and the magnetic field, as plotted in Fig.~\ref{fig:Fig3}(a). Resonator transmission and phase are simultaneously fitted to input-output theory and the qubit-photon coupling, qubit tunnel coupling and qubit linewidth are extracted. Descriptions to the method and formulas can be found in the main text and methods section.}

\corr{We observe a qualitatively similar curve shape to device A in the main text. Again,} we model the DQD by an effective two electron Hamiltonian which allows us to fit the gate voltage and field dependence of the IDT (white dashed line in Fig.~\ref{fig:Fig3}(a)).
We find that the magneto-dispersion of the IDT is well fitted using the following fit parameters namely the spin-conserving singlet and triplet tunnel rates $t_c^S/2\pi\approx29$\,GHz, and
$t_c^T/2\pi\approx37$\,GHz, the singlet-triplet coupling rate
$t_{\rm{SO}}/2\pi\approx 5$\,GHz, the electron g-factors of the right and left dots, $g_R\approx 1.8$ and $ g_L\approx 2.8$, as well as the singlet-triplet energy splitting $\Delta_{\rm{ST}}/2\pi\approx 79$\,GHz.
As for device A, these fit parameters are consistent with parameters obtained previously in this material system~\cite{fasth2007direct,nadj2010disentangling,nilsson2018tuning,trif2008spin,junger2020magnetic}.

As described \corr{in the main text}, we extract the the qubit tunnel amplitude $t$, the qubit linewidth $\gamma$, and the qubit-photon coupling strength $g$ as a function of $B$, which we plot in Figure~\ref{fig:Fig3}(b). \corr{A notable difference to device A, is the higher tunnel rate of device B. Unlike device A, device B has a qubit frequency predominantly above the resonator frequency and therefore anti-crosses only in small regions of the dispersion.}

\corr{We will now shortly discuss the different regimes of the qubit, analogue to device A.
Fig.~\ref{fig:Fig3}(c) shows the corresponding DQD level structure based on the fit parameters as a function of $\epsilon$ for different magnetic field.}

\corr{At a low magnetic fields around $B=0.1$\,T, we observe again a singlet charge qubit with Zeeman-split triplets in the weak coupling limit. Again, we observe the characteristic double-dip structure between $B\sim 0.03$\,T and $B\sim0.3$\,T of an even IDT.}

\corr{As shown in the second panel of Fig.~\ref{fig:Fig3}(c) at high enough field $T_{1,1}^+$ becomes the ground state for $\epsilon<0$.
The spin-orbit interaction couples the singlet and triplet states, leading to an anti-crossing between $S_{2,0}$ and $T_{1,1}^+$.}

Consistent with the interpretation of the formation of a singlet-triplet qubit, we measure an approximately constant tunneling rate $t$ between $B\sim0.5$\,T and $B\sim1.1$\,T.
We extract the average spin-orbit tunneling to be $\bar{t}_{\rm{so}}=4.0\pm 0.3$\,GHz. % where the error bar is the root variance.
At $B\approx 1.3$\,T, $\chi$ becomes positive. This is interpreted as a drop of the tunnel rate below the resonator frequency, $2t<\omega_0$.
This decline in $t$ is not captured by our simplified Hamiltonian model and we speculate that changes in the orbital structure of a many-electron DQD could be the reason.

At a magnetic field of $B\approx 1.7$\,T, we observe a resonant interaction between the resonator and the singlet-triplet qubit leading to the anti-crossing as discussed in section \ref{sec:anti-crossing}.
As seen in the level structure in Fig.~\ref{fig:Fig3}(c) at $B=1.65$\,T, the triplet state $T_{2,0}^+$ becomes relevant.
This results in a level repulsion between $T_{2,0}^+$ and $T_{1,1}^+$ and hence leads to a reduced splitting between the $S_{2,0}$ level and the $T_{1,1}^+$ states.
In Fig.~\ref{fig:Fig3}(c), this is illustrated by the smaller level gap (black arrow) compared to the one at $B=0.7$\,T.
 
The level structure at very large magnetic fields is plotted at $B\approx2$\,T in the right panel of Fig.~\ref{fig:Fig3}(c). \corr{Here, we observe the triplet charge qubit.}
%, forming a charge qubit with triplet spin character.
As discussed in the main text, the triplet charge qubit has a larger frequency detuning from the resonator frequency than the singlet-triplet qubit, leading to a smaller resonator shift.
\section*{Bibliography}
\bibliography{ref_supp}